\documentclass[preprint]{aastex}
\usepackage{amsmath}
\usepackage{placeins}

\begin{document}

\title{Peering through the Dust: NuSTAR Observations of Two FIRST-2MASS Red Quasars
}

\author{Stephanie M. LaMassa$^{1,2,3}$, 
Angelo Ricarte$^{4}$,
Eilat Glikman$^5$, 
C. Megan Urry$^{1,2}$,
Daniel Stern$^6$,
Tahir Yaqoob$^7$,
George B. Lansbury$^8$,
Francesca Civano$^{1,2,9,10}$,
Steve E. Boggs$^{11}$,
W. N. Brandt$^{12,13,14}$,
Chien-Ting J. Chen$^{12,13}$,
Finn E. Christensen$^{15}$,
William W. Craig$^{11,16}$,
Chuck J. Hailey$^{17}$,
Fiona Harrison$^{18}$,
Ryan C. Hickox$^{9}$,
Michael Koss$^{19}$,
Claudio Ricci$^{20}$,
Ezequiel Treister$^{21}$,
Will Zhang$^3$
}

\affil{$^1$Yale Center for Astronomy \& Astrophysics, Physics Department, P.O. Box 208120, New Haven, CT 06520, USA;,
$^2$ Department of Physics, Yale University, P.O. Box 208121, New Haven, CT 06520, USA,
$^3$ NASA Goddard Space Flight Center, Greenbelt, MD 20771, USA,
$^4$Department of Astronomy, Yale University, New Haven, CT 06511, USA,
$^5$Middlebury College, Department of Physics, Middlebury, VT 05753, USA,
$^6$Jet Propulsion Laboratory, California Institute of Technology, 4800 Oak Grove Drive, Mail Stop 169-221, Pasadena, CA 91109, USA,
$^7$Department of Physics, University of Maryland Baltimore County, 1000 Hilltop Circle, Baltimore, MD 21250
$^8$Center for Extragalactic Astronomy, Department of Physics, University of Durham, South Road, Durham DH1 3LE, UK,
$^9$Department of Physics and Astronomy, Dartmouth College, 6127 Wilder Laboratory, Hanover, NH 03755,USA,
$^{10}$Harvard-Smithsonian Center for Astrophysics, 60 Garden Street, Cambridge, MA 02138, USA, 
$^{11}$Space Sciences Laboratory, University of California, Berkeley, CA 94720, USA,
$^{12}$Department of Astronomy and Astrophysics, The Pennsylvania State University, 525 Davey Lab, University Park, PA 16802, USA,
$^{13}$Institute for Gravitation and the Cosmos, The Pennsylvania State University, 525 Davey Lab, University Park, PA, 16802, USA,
$^{14}$Department of Physics, The Pennsylvania State University, 525 Davey Lab, University Park, PA 16802, USA,
$^{15}$DTU Space—National Space Institute, Technical University of Denmark, Elektrovej 327, DK-2800 Lyngby, Denmark,
$^{16}$Lawrence Livermore National Laboratory, Livermore, CA 945503, USA,
$^{17}$Columbia Astrophysics Laboratory, Columbia University, NY 10027, USA,
$^{18}$Cahill Center for Astronomy and Astrophysics, California Institute of Technology, 1216 E. California Blvd, Pasadena, CA 91125, USA,
$^{19}$Institute for Astronomy, Department of Physics, ETH Zurich, Wolfgang-Pauli-Strasse 27, CH-8093 Zurich, Switzerland,
$^{20}$Instituto de Astrof\'isica,Pontificia Universidad Cat\'olica de Chile,Vicu$\tilde{n}$a Mackenna 4860, 7820436 Macul, Santiago, Chile,
$^{21}$Universidad de Concepci\'on, Departamento de Astronom\'ia, Casilla 160-C, Concepci\'on, Chile
}

\begin{abstract}

\keywords{galaxies:active, infrared:galaxies, quasars: F2M0380+3759, F2M1227+3214; X-rays: F2M0830+3759, F2M1227+3214 }

Some reddened quasars appear to be transitional objects in the merger-induced black hole growth/galaxy evolution paradigm, where a heavily obscured nucleus starts to be unveiled by powerful quasar winds evacuating the surrounding cocoon of dust and gas. Hard X-ray observations are able to peer through this gas and dust, revealing the properties of circumnuclear obscuration. Here, we present {\it NuSTAR} and {\it XMM-Newton}/{\it Chandra} observations of FIRST-2MASS selected red quasars F2M 0830+3759 and F2M 1227+3214. We find that though F2M 0830+3759 is moderately obscured ($N_{\rm H,Z} = 2.1\pm0.2 \times10^{22}$ cm$^{-2}$) and F2M 1227+3214 is mildly absorbed ($N_{\rm H,Z} = 3.4^{+0.8}_{-0.7}\times10^{21}$ cm$^{-2}$) along the line-of-sight, heavier global obscuration may be present in both sources, with $N_{\rm H,S} = 3.7^{+4.1}_{-2.6} \times 10^{23}$ cm$^{-2}$ and  $< 5.5\times10^{23}$ cm$^{-2}$, for F2M 0830+3759 and F2M 1227+3214, respectively. F2M 0830+3759 also has an excess of soft X-ray emission below 1 keV which is well accommodated by a model where 7\% of the intrinsic AGN X-ray emission is scattered into the line-of-sight. While F2M 1227+3214 has a dust-to-gas ratio ($E(B-V)$/$N_{\rm H}$) consistent with the Galactic value, the $E(B-V)$/$N_{\rm H}$  value for F2M 0830+3759 is lower than the Galactic standard, consistent with the paradigm that the dust resides on galactic scales while the X-ray reprocessing gas originates within the dust-sublimation zone of the broad-line-region. The X-ray and 6.1$\mu$m luminosities of these red quasars are consistent with the empirical relations derived for high-luminosity, unobscured quasars, extending the parameter space of obscured AGN previously observed by {\it NuSTAR} to higher luminosities.
\end{abstract}

\section{Introduction}

Supermassive black holes (SMBHs) reside at the center of almost every massive galaxy.  There, these objects can grow by accretion and be observed as active galactic nuclei (AGN). In addition to being among the most energetic sources in the Universe, AGN may also play a key role in the evolution of the galaxies in which they live \citep[e.g.,][]{kormendyho,ah,heckman}. SMBH mass is correlated with galaxy mass \citep[e.g.,][]{ferrarese, gebhardt, graham}, suggesting common physical processes that link the life cycles of both systems. Additionally, theoretical simulations invoke feedback from thermal and/or kinetic energy associated with black hole accretion to match observed galaxy properties \citep[e.g.,][]{scannapieco}. One mechanism that triggers SMBH fueling and concurrent galaxy growth is major galaxy mergers, where the nucleus is predicted to be enshrouded by large amounts of dust and gas before powerful AGN winds expel this obscuring material \citep[e.g.,][]{sanders,hopkins2005}, unveiling the typical unobscured AGN identified by optical surveys such as the Sloan Digital Sky Survey \citep[SDSS,][]{york,schneider}. However, the phase where AGN feedback heats gas within the host galaxy, thereby regulating star formation, is expected to be short-lived, on the order of several hundred million years \citep[e.g.,][]{hopkins2008,glikman2012}, meaning that such objects are rare. 

Though these transitional AGN occupy a low space density in the Universe \citep[e.g., $1.2\pm0.1\times 10^{-3}$ deg$^{-2}$ to a $K$ = 14.5 magnitude (AB) limit;][]{glikman2012}, they provide a unique opportunity to study SMBH and galaxy co-evolution in action. Among the best candidates for these systems are ``red quasars'' which are a class of obscured AGN quite different from the type described by the unification model \citep{antonucci,urry,netzer}. Unlike the Type 2 AGN explained by unification, which are identified by having only narrow emission lines in their optical spectra, red quasars, as defined by \citet{glikman2004}, have broad emission lines akin to Type 1 to Type 1.9 AGN. However, they have large amounts of dust that attenuate optical emission and reddens their spectra, making them difficult to identify based on optical-only diagnostics. These AGN are thus discovered by their red optical to infrared colors \citep[i.e., either $R-K>4-5$, $B-K>6.5$, {$R-[3.6]>4$}, {$R-[4.5]>6.1$}, or $F$(24$\mu$m)/$F(R)>1000$;][]{glikman2004,glikman2007,brusa2005,brusa2010,hickox,fiore2008,fiore2009} and/or red infrared colors \citep[i.e., $J-K > 1.7-2.5$, $W1 - W2 > 0.8$\footnote{$W1$ and $W2$ refer to the {\it WISE} passbands at 3.4$\mu$m and 4.6$\mu$m, respectively. We note that such color selections described above do not exlusively identify broad-lined AGN, but also select narrow-lined AGN or those lacking any emission lines.};][]{glikman2004,banerji2012,stern2012,assef}. {\it WISE} has also identified a population of ``hot, dust-obscured galaxies'' (Hot DOGs), which, with infrared luminosities exceeding 10$^{13}$ $L_{\odot}$, may be the most luminous AGN in the Universe \citep{assef2015,tsai}.\footnote{{\it WISE} Hot DOGs are selected by having very red {\it WISE} colors, i.e. strong detections in the $W3$ (12$\mu$m) and $W4$ (22$\mu$m) bands with faint or non-detections in $W1$ and $W2$. These sources include both narrow-lined and broad-lined AGN.}

Additional multi-wavelength constraints, such as detections in the radio \citep{glikman2004,glikman2007,glikman2012,glikman2013} and high X-ray to optical fluxes \citep{brusa2010}, are sometimes invoked in identifying red quasar candidates to mitigate contamination from dusty star-forming galaxies or stars. Such multi-wavelength diagnostics have revealed sources which appear to be in a transitional stage between galaxy coalesence and evacuation.  For instance, galactic-scale outflows have been detected in red quasars initially discovered on the basis of their X-ray, optical, and infrared properties \citep{brusa2007,brusa2015}. Radio observations of 2MASS-selected reddened quasars have revealed young radio jets, suggesting that they are in the early stages of black hole growth where the expansion of radio lobes can impart feedback onto the host galaxy \citep{antonis2012}. Optical to far-infrared photometry of a sample of reddened quasars show evidence of outflows that can inhibit host-galaxy star formation \citep{farrah}. {\it Hubble} imaging of red quasars presented by \citet{glikman2004}, found by cross-correlating the FIRST and 2MASS surveys, shows that they have ``train-wreck'' morphological traits indicative of merger activity \citep{urrutia2008,glikman2015}. After correcting for extinction, these FIRST-2MASS red quasars are among the most luminous AGN at every redshift \citep{glikman2012,banerji2015}, similar to the Hot DOGs discovered by {\it WISE} \citep{assef2015,tsai}.

Though reddened AGN have been identified in X-ray surveys \citep[e.g.,][]{hickox,fiore2008,fiore2009,brusa2010}, there are few studies of targeted X-ray follow-up of infrared-selected red quasar samples \citep[e.g.,][]{brusa2005}. While most of these targeted red quasars are detected in X-rays, very few have an adequate number of observed photons to enable characterization of their X-ray properties. For instance, \citet{wilkes2002} studied a sample of 26 2MASS-selected reddened AGN with {\it Chandra}, but due to the short exposure times (1 - 4.5 ks), analysis of the X-ray obscuration was limited to hardness ratios\footnote{$HR \equiv  (H-S)/(H+S)$, where $H$ represents the number of counts in the hard band and $S$ is the number of counts in the soft band, which in \citet{wilkes2002} are the 2.5 - 8 keV and 0.5  - 2 keV bands, respectively.} since insufficient counts were detected for a proper spectral-fitting analysis. \cite{wilkes2005} followed-up five of these AGN with {\it XMM-Newton}, two of which were narrow-lined objects and three of which were broad-lined AGN, and detected hundreds to over a thousand counts, enabling spectral fitting that better reveal their X-ray properties. They found three objects to have moderate X-ray absorption ($N_{\rm H} \sim 10^{22}$ cm$^{-2}$), as well as a ``soft excess'' component below 2 keV in the three broad-lined AGN, which they attributed to emission from extended ionized gas.

Unlike the 2MASS-selected reddened AGN, which tend to be lower luminosity sources at a median redshift of $\sim$0.23 and include many Type 2 objects \citep{cutri}, FIRST-2MASS red quasars are Type 1 AGN at much higher luminosities \citep{glikman2004,glikman2007,brusa2005,banerji2012,banerji2015}, and there is mounting evidence that their reddening is largely due to host galaxy dust related to merger activity \citep[e.g.,][]{urrutia2008,glikman2015}. \citet{urrutia} presented the first X-ray analysis of FIRST-2MASS red quasars. Similar to the study from \citet{wilkes2002}, 11 of the 12 objects targeted by {\it Chandra} were detected, but only six had enough counts for a crude spectral fit. One of these objects, F2M 0830+3759, was followed-up with {\it XMM-Newton} for $\sim$50 ks by \citet{piconcelli}, where they obtained a significantly flatter spectral index than reported by \citet{urrutia} ($\Gamma = 1.51\pm0.06$ versus $\Gamma=2.9\pm0.1$), indicating that the properties derived from the shorter X-ray exposures may reflect limited signal-to-noise in the spectra rather than trace intrinsic physical processes of the quasars. Interestingly, \citet{piconcelli} report a soft-excess below 1 keV, similar to what is reported by \citet{wilkes2005} in their three broad-lined AGN observed by {\it XMM-Newton}. 

Since these studies have been published, substantial improvements in tools used for the X-ray modeling of obscured AGN have become available. In particular, the \citet{mytorus} MYTorus model, the \citet{ikeda} torus model, the spherical and toroidal absorption models of \citet[][BNTorus]{bn_torus}, and the \citet{liu2014} clumpy torus model self-consisently account for the transmitted, Compton-scattered, and Fe fluorescent line emission through an obscuring medium with column densities ranging from moderate ($N_{\rm H} = 10^{22}$ cm$^{-2}$) to extremely Compton-thick ($N_{\rm H} = 10^{26}$ cm$^{-2}$ for the BNTorus model, and $N_{\rm H} = 10^{25}$ cm$^{-2}$ for the remaining models). However, only the BNTorus and MYTorus models are publicly available. The latter model has the capability to emulate a patchy obscuring medium, where the line-of-sight and global column densities are independent parameters. This mode may be of particular relevance to red quasars where presumably the accretion disk is viewed directly (allowing broad lines in the optical and/or infrared spectra to be observed), while large amounts of obscuring gas may be present out of the line-of-sight \citep[see Figure 2 in][]{yaqoob2012}. Such physically motivated models allow a more reliable estimate of the gas column density to be calculated, as well as the obscuring geometry to be constrained, compared to the phenomenological absorbed power law models used in previous studies.

Here, we use {\it NuSTAR} \citep{harrison} and archival {\it Chandra} and {\it XMM-Newton} observations to study the broad-band X-ray properties of two FIRST-2MASS selected red quasars from the sample reported in \citet{glikman2012}: F2M 0830+3759 and F2M 1227+3214. Of the fourteen FIRST-2MASS red quasars observed to date in X-rays \citep{urrutia,evans}, these 2 sources are the nearest and have X-ray count rates that indicated they would be bright enough to ensure detection of $\simeq 1000$ counts in the {\it NuSTAR} bandpass with a relatively short exposure time. We calculated the $\alpha_{\rm  IX}$\footnote{$\alpha_{\rm IX} = \frac{ {\rm Log} (\nu L_{\rm 2keV}/(\nu L_{\rm 12\mu m})} {{\rm Log}(\nu_{\rm 2keV}/\nu_{12 \mu m})}$, where $\nu L_{\rm 2keV}$ and $\nu L_{\rm 12\mu m}$ are the monochromatic luminosites (erg s$^{-1}$ Hz$^{-1}$)  at 2 keV and 12$\mu$m, respectively, in the rest-frame \citep[e.g.][]{gandhi}.} values for the FIRST-2MASS sources that have existing X-ray information, assuming a powerlaw spectrum where $\Gamma=1.8$ for all sources to derive the monochromatic 2 keV luminosity, and calculated reddening-corrected 12$\mu$m monochromatic luminosities from the optical and infrared spectra. F2M 0830+3759 and F2M 1227+3214 have the highest $\alpha_{\rm ix}$ values, indicating that they are stronger in X-rays compared to the other FIRST-2MASS sources yet studied.

We modeled the X-ray spectra for F2M 0830+3759 and F2M 1227+3214 over the energy range 0.5-79 keV, where the high energy coverage of {\it NuSTAR} is essential for obtaining the best constraints on the X-ray obscuration, the intrinsic X-ray continuum, and the geometry of the X-ray reprocessor. In addition to assessing the X-ray obscuring medium and how it relates to the optical reddening, we compare the observed X-ray luminosities with the infrared 6.1$\mu$m luminosities, placing these objects in context with other obscured AGN studied by {\it NuSTAR}. Throughout, we adopt a cosmology of  $H_{0}$ = 70 km s$^{-1}$ Mpc$^{-1}$, $\Omega_{\rm M} = 0.3$ and $\Omega_{\Lambda}= 0.7$, and use AB magnitudes.

\section{Multi-wavelength Observations}

\subsection{Optical and Infrared Properties of F2M 0830+3759 and F2M 1227+3214}
F2M 0830+3759 and F2M1227+3214 were selected by the FIRST-2MASS survey, which required them to be detected independently in the FIRST and 2MASS surveys, within a 2$^{\prime\prime}$ search radius, effectively restricting the sample to sources with strong core radio emission. Objects also had to pass optical to near-infrared colors cuts of $R-K>4$ and $J-K>1.7$, where $J$ and $K$ come from the 2MASS survey \citep{skrutskie} while the $R$ magnitude comes from the Guide Star Catalog II \citep[GSCII;][]{lasker} which is produced by digitizing the second-generation Palomar Observatory and UK Schmidt sky surveys \citep[POSS-II;][]{reid}.  The resultant sample contained 120 quasars with $E(B-V)\ge 0.1$, reaching reddenings as high as $E(B-V) = 1.5$. The sample's redshift range extends from $z=0.14$ to $z=3.05$.

Both F2M 0830+3759 and F2M 1227+3214 appear in the spectral atlas of \citet{glikman2012} with optical spectra from the Keck and Lick observatories, respectively, dating back to 2001 and 1998, respectively.  Since then, F2M 1227+3214 had been observed by SDSS with a spectrum that extends to longer wavelength. We obtained a near-infrared spectrum of this source on UT 2015 March 13 at the Apache Point Observatory 3.5m telescope with the TripleSpec cross-dispersed near-infrared spectrograph \citep{wilson}.  The quasar was exposed for 16 minutes using an ABBA dither pattern along the slit followed by an observation of an A0V telluric standard. The data were reduced using the Spextool software following the procedude described in \citet{cushing} and \citet{vacca}.

\begin{figure}[ht]
\centering
\includegraphics[scale=0.45]{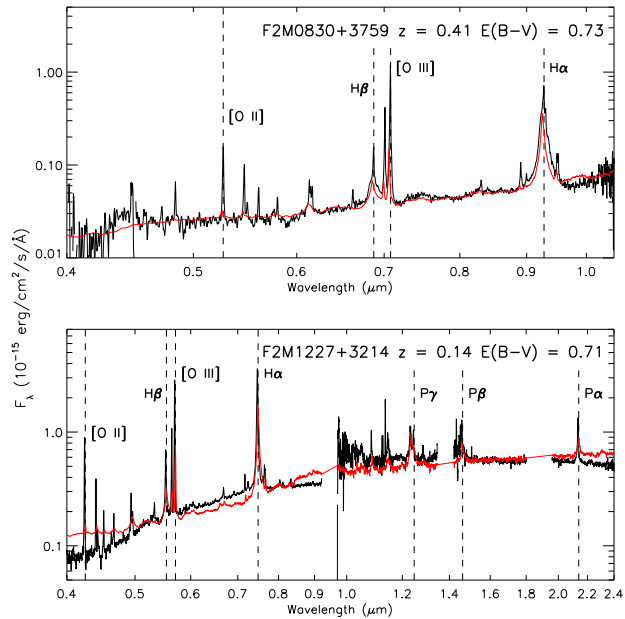}
\caption{\label{rq_ir_opt_spec} {\it Top}: Keck optical spectrum of F2M 0830+3759. {\it Bottom}: SDSS \& APO TripleSpec spectra of F2M 1227+3214. In both sets of spectra, prominent AGN emission lines are marked and labeled. The red lines show the best-fit reddened optical-to-near-infrared template of \citet{glikman2006} applied to the spectra. The poorer fit of the template to F2M 1227+3214 suggests that the derived reddening value is a lower-limit.}
 \end{figure}

We plot the spectra of both sources in Figure \ref{rq_ir_opt_spec}, showing the optical spectrum of F2M 0830+3759 in the top panel, and the newly combined optical-through-near-infrared spectrum of F2M 1227+3214 in the lower panel.  We plot both spectra on a log-log scale to enhance the appearance of features over the broad wavelength and flux ranges apparent in these spectra. Vertical dashed lines mark the location of prominent AGN lines, namely Pa$\alpha$, Pa$\beta$, Pa$\gamma$, H$\alpha$, [\ion{O}{3}] 5007\AA, H$\beta$ and [\ion{O}{2}] 3727\AA. Both quasars also show strong emission from [\ion{Ne}{3}] at 3869\AA\ and 3967\AA. 

To determine the reddening of these sources, we fit the optical-to-near-infrared quasar template of \citet{glikman2006} attenuated by a Small Magellanic Cloud (SMC) dust law following the procedure described in Section 5 of \citet{glikman2012}.  Figure \ref{rq_ir_opt_spec} shows the best-fit reddened template plotted atop the data (red line). We recover the same $E(B-V)=0.73$ value for F2M 0830+3759 as reported in \citet{glikman2012}, which appears well-fit by this model. However, the added near-infrared spectrum of F2M 1227+3214 lowers our measured $E(B-V)$ from the \citet{glikman2012} value of 0.94 mag to 0.71 mag.  The reddening law produces a poorer fit to this system, especially at the shortest wavelengths, suggesting that the new reddening value may be an underestimate \citep[see][for a discussion of different reddening laws applied to F2M red quasars, where the SMC law returns the best fit for these sources]{glikman2012}.  

\begin{figure}[ht]
\centering
\includegraphics[scale=0.45]{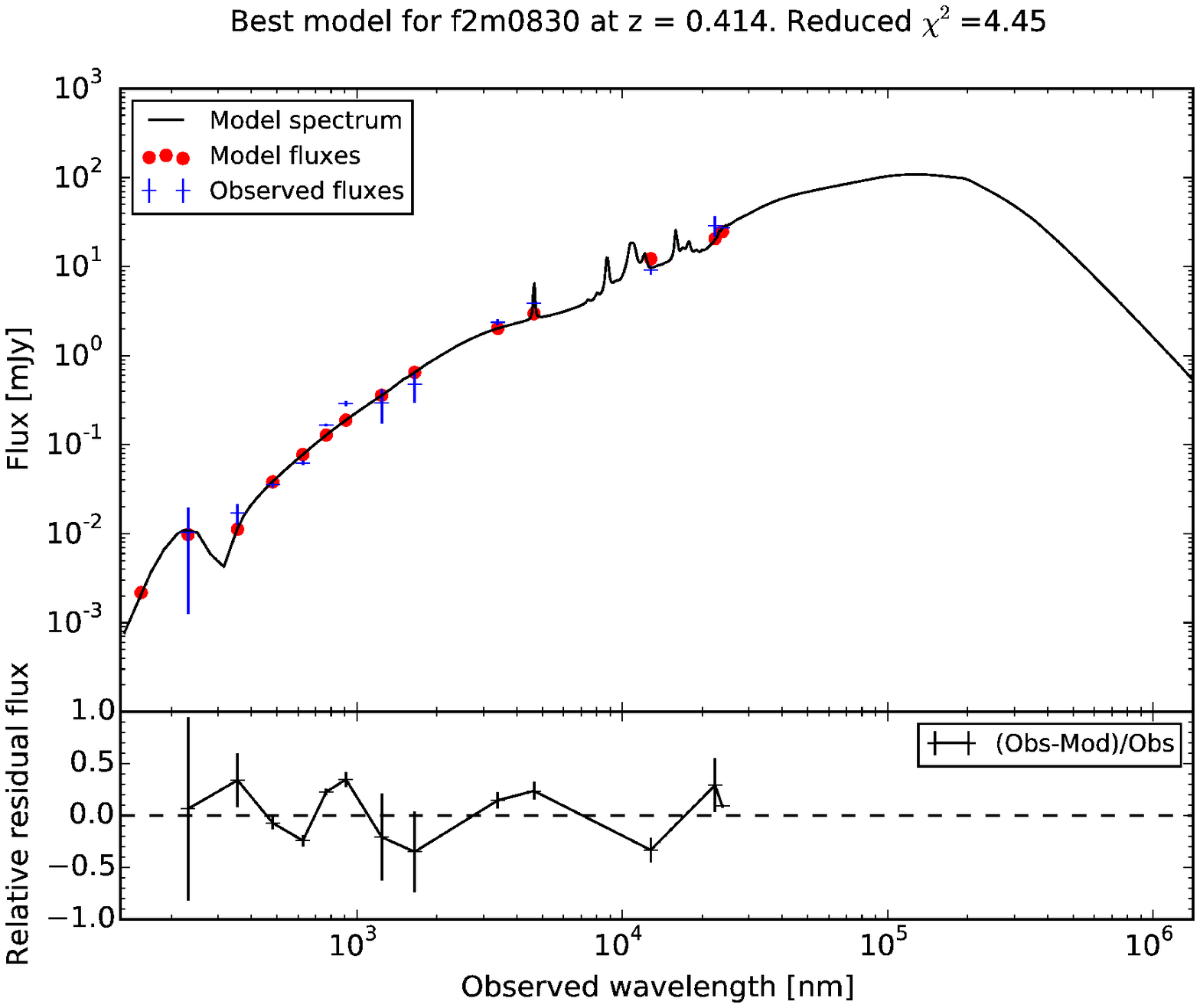}
\includegraphics[scale=0.45]{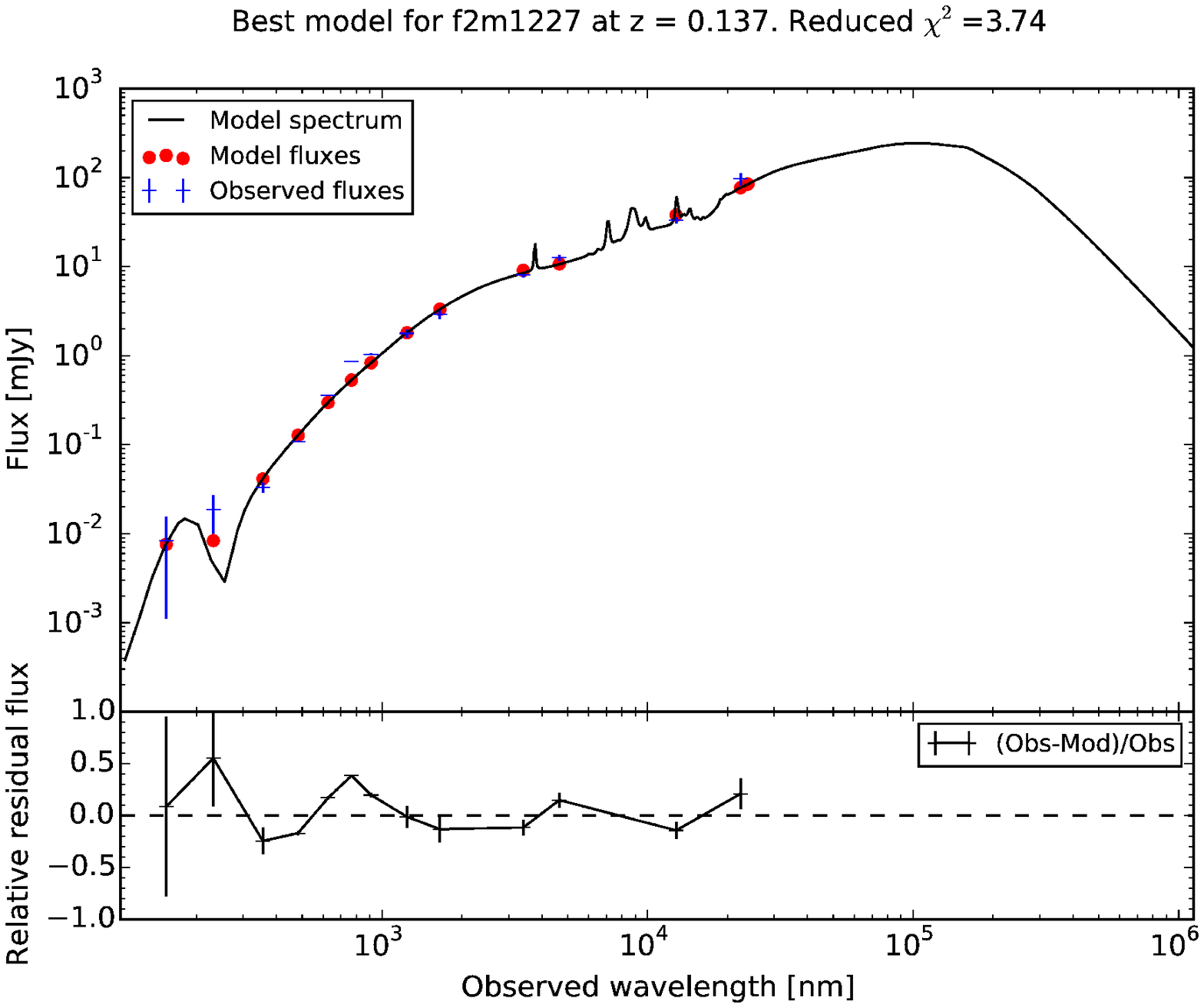}
\caption{\label{rq_seds} Rest-frame SEDs of ({\it top}) F2M 0830+3759 and ({\it bottom}) F2M 1227+3214 which were fitted using CIGALE \citep{noll,serra}. The model includes a star formation history with two decreasing exponential functions, re-processed dust emission from the \citet{draine} model, and AGN dust emission as parameterized by the \citet{fritz} model. }
 \end{figure}

In addition to the optical and near-infrared (in the case of F2M 1227+3214) spectroscopy, we also utilize photometric data from the optical, via SDSS, through the near-infrared, via 2MASS, to the mid-infrared, via {\em WISE}  \citep[F2M 0830+3759 was also observed with {\em Spitzer} and analyzed independently by][]{urrutia2012}.  These data provide at least a dozen photometric data points which we use to model the spectral energy distribution (SED) using the Cigale SED fitting code \citep{noll,serra}. Figure \ref{rq_seds} shows the resultant SED fits at rest-frame wavelengths with photometric data overplotted.  The SED model includes components for a star formation history with a double decreasing exponential function, \citet{draine} dust emission, and \citet{fritz} AGN-heated dust emission.

To estimate the bolometric luminosities of these quasars, we integrate under the model between 3500\AA\ and 400$\mu$m; reddening attenuates the ultraviolet emission at lower wavelengths which is then reprocessed into infrared emission which we measure here.  We find luminosities of $L_{\rm bol} = 8.5\times10^{45}$ erg s$^{-1}$ for F2M 0830+3759 and $L_{\rm bol} = 2.3\times10^{45}$ erg s$^{-1}$ for F2M 1227+3214.  Since most of the energy is attributed to the dust emission at wavelengths beyond $\sim 20\mu$m, where we do not have data to constrain the model, we also compute a conservative lower limit to the luminosities of these quasars by integrating only out to $24\mu$m.  Our conservative limits for the luminosities are $L_{\rm bol} = 3.9\times10^{45}$ erg s$^{-1}$ for F2M 0830+3759 and $L_{\rm bol} = 1.3\times10^{45}$ erg s$^{-1}$ for F2M 1227+3214. We summarize the reddening values and bolometric luminosities calculated here, as well as the rest-frame, non-absorption corrected 6.1$\mu$m luminosities ($L_{\rm 6.1\mu m}$) derived from the SED modeling in Table \ref{qso_prop}.

\begin{table}[h]
\small
   \begin{center}
    \caption{\label{qso_prop}Red Quasar Properties}
   \begin{tabular}{llllllll} 
   \hline
   \hline
   Source & R.A. & Decl. & {\it z} & $E(B-V)$ & Log ($L_{\rm 1.4 GHz}$) & Log ($\nu L_{\rm 6.1\mu m}$) & Log ($L_{\rm bol}$)$^{1}$\\
          &           &           &           &                & (erg s$^{-1}$ Hz$^{-1}$) & (erg s$^{-1}$)                    & (erg s$^{-1}$) \\
   \hline
   F2M 0830+3759 & 08 30 11.12 & +37 59 51.8 & 0.41 & 0.73 & 31.60 & 45.10 & 45.59 \\
   F2M 1227+3214 & 12 27 49.15 & +32 14 59.0 & 0.14 & 0.71 & 30.53 & 44.60 & 45.12 \\
\hline
\hline
\multicolumn{7}{l}{$^1$Bolometric luminosities are calculated from integrating the model SED fit between 3500\AA\ and 24$\mu$m, }\\
\multicolumn{7}{l}{i.e., corresponding to the conservative value quoted in the text.}
   \end{tabular}
   \end{center}
\end{table}

\begin{table}[h]
   \begin{center}
    \caption{\label{obs_sum}Summary of X-ray Observations}
   \begin{tabular}{llllll} 
   \hline
   \hline
   Source & \multicolumn{2}{c}{{\it NuSTAR} Observations} &  \multicolumn{3}{c}{Archival Observations} \\
              & ObsID & Date & Observatory & ObsID & Date \\
    \hline

   F2M 0830+3759 & 60001109002 & 2014 Sep & {\it XMM-Newton} &  0554540201 & 2008 Nov \\
   F2M 1227+3214 & 60001108002 & 2014 Jul & {\it Chandra} &  4138 &  2003 Apr \\
\hline
\hline
   \end{tabular}
   \end{center}
\end{table}

\subsection{Radio Properties}
Though these quasars were selected from the FIRST survey, they are not necessarily radio loud objects. Before calculating radio loudness ($R_{m}$ = log($F_{\rm radio}$/$F_{\rm optical}$)), the optical emission in $g$-band is corrected for extinction using the measured $E(B-V)$ values. Following \citet{ivezic}, \citet{glikman2007} calculated radio loudness using:
\begin{equation}
R_{m} = 0.4(g_{\rm corr} - t),
\end{equation}
where $g_{\rm corr}$ is the extinction-corrected $g$-band magnitude and $t$ is the FIRST flux density ($t = 2.5$log($F_{\rm int}$/3631 Jy)). We find that $R_{m}$ = 1.14 and 0.7 for F2M 0830+3759 and F2M 1227+3214, respectively. While the latter source is radio quiet ($R_{m}<1$), F2M 0830+3759 is considered radio-intermediate \citep[$1 < R_{m} < 2$;][]{miller}. \citet{miller} demonstrated that radio-intermediate quasars often have an excess of X-ray emission compared to radio-quiet quasars, with the amount of excess ranging from slight to as high as a factor of several. This enhanced X-ray brightness, putatively from jet-linked emission, becomes more pronounced on average for the radio-loud quasar population. 

VLA data exist for F2M 0830+3759, which has a measured radio spectral index ($\alpha_{\rm 1.4 GHz/8.3 GHz}$, $S_{\nu} \propto \nu^{\alpha}$) of -1.06 \citep{glikman2007}. This is steeper than observed in flat-spectrum objects, where $\alpha_{\rm 1.4 GHz/8.3 GHz}>-0.5$, which are interpreted as having radio-jets beamed in the direction along the line-of-sight. We can therefore assume that any jet-associated emission does not directly intersect our line-of-sight. However, as F2M 0830+3759 and F2M 1227+3214 are stronger in X-rays compared with the red quasars observed thus far, there can be a boost to the X-ray emission from a jet-linked contribution.

\FloatBarrier

\subsection{{\it NuSTAR}}
{\it NuSTAR}, launched in 2012 June, is the first focusing hard X-ray telescope above 10 keV in orbit, sensitive to energies between 3 to 79 keV \citep{harrison}. It consists of two co-aligned mirror modules which focus hard X-rays onto two Focal Plane Modules, FPMA and FPMB. With a field-of-view of $\approx12^{\prime} \times 12^{\prime}$, {\it NuSTAR} has an angular resolution of 18$^{\prime\prime}$ (FWHM). Due to its high-energy sensitivity, it is an ideal instrument for studying obscured AGN as it recovers X-ray emission that is attenuated at lower energies.

The details of the X-ray observations for F2M 0830+3759 and F2M 1227+3214 are summarized in Table \ref{obs_sum}. F2M 0830+3759 was observed with {\it NuSTAR} for 22 ks on UT 2014 September 19 (ObsID: 60001109002), while F2M 1227+3214 was observed for 23 ks on UT 2014 July 31 (ObsID: 60001108002).  Data were reduced with {\it nupipeline}, which is part of the {\it NuSTAR} Data Analysis Software NuSTARDAS v.1.4.1, CALDB v.20140814 (Perri et al. 2013).\footnote{http://heasarc.gsfc.nasa.gov/docs/nustar/analysis/nustar\_swguide.pdf} The spectra were extracted from a circular region 40$^{\prime\prime}$ in radius around the source from both the FPMA and FPMB detectors.  This radius was chosen to be large enough to encompass emission from the quasar whilst being small enough to minimize background photons. The background spectra were extracted from annuli with inner radii of 90$^{\prime\prime}$ and outer radii of 240$^{\prime\prime}$, centered on the quasar; no serendipitous sources are detected in these background regions.  These spectra were grouped by 20 counts per bin, with
1169$\pm$34 net counts detected in F2M 0830+3759 and 1082$\pm$33 net counts detected in F2M 1227+3214. We note that as neither of these sources were detected by {\it Swift} BAT \citep{baumgartner}, these data represent the first observations of these quasars above 10 keV.

\subsection{{\it XMM-Newton}: F2M 0830+3759}
{\it XMM-Newton} observed F2M 0830+3759 for 52 ks in 2008 November \citep[PI: Piconcelli, ObsID: 0554540201;][]{piconcelli}. Though it was also observed with {\it Chandra} for 9 ks \citep{urrutia}, we use only the {\it XMM-Newton} data due to the superior signal-to-noise from the longer observation. The data were reduced with the {\it XMM-Newton} Science Analysis System (SAS) package using HEASOFT v.6.16 to apply standard filtering to the events file and remove time intervals with background flaring. We extracted spectra from a 35$^{\prime\prime}$ aperture radius centered on the source for all three {\it XMM-Newton} detectors (PN, MOS1, and MOS2). For the MOS observations, we extracted the background region from an annulus around the source, free of serendipitous sources, with an inner radius of 45$^{\prime\prime}$ and outer radius of 100$^{\prime\prime}$. Since the object was close to the chip gap in the PN dector, the background here was instead extracted from several source-free circular regions near the quasar. About 7000 net counts were detected by PN and $\sim$2700 counts by each of the MOS detectors. The MOS spectra were grouped such that each bin contains at least 20 counts, while the superior sensitivity of PN allowed us to group the data such that each bin contains at least 50 counts.

\subsection{{\it Chandra}: F2M 1227+3214}

F2M 1227+3214 was targeted with {\it Chandra} ACIS-I \citep{garmire} on 2003 April 30 for 3.7 ks (PI: Laurent-Muehleisen, ObsID: 4183), though this work represents the first time results from these data are published other than in the {\it Chandra} Source Catalog \citep{evans}. We processed the data with the CIAO v4.5, with CALDB v4.5.5.1 \citep{fruscione}, using the {\it chandra\_repro} task to produce a filtered events file, removing periods of anomalously high background. Due to {\it Chandra}'s superior angular resolution, the source spectrum was extracted using a 5$^{\prime\prime}$ radius aperture around the object using the CIAO tool {\it specextract}, with the background extracted from an annulus around the quasar with inner radius 10$^{\prime\prime}$ and outer radius 30$^{\prime\prime}$. The spectrum was grouped by a minimum of 15 counts per bin, with a total of 834$\pm$29 net counts detected.

\section{Spectral Analysis}
\label{sec:spectralAnalysis}
We simultaneously fit the {\it NuSTAR} spectra (3-79 keV) and the archival X-ray spectra (0.5-8 keV for {\it Chandra}, 0.5-10 keV for {\it XMM-Newton}) using XSpec v12.8.2 \citep{arnaud}, where the background is automatically subtracted. A constant factor was included in the modeling to account for calibration differences between {\it NuSTAR} and {\it Chandra}, and {\it NuSTAR} and {\it XMM-Newton}. For F2M 0830+3759, we find that the ratio between the FPMA (FPMB) normalization and the {\it XMM-Newton} PN detector is $1.26\pm0.12$ ($1.34\pm0.13$), which is higher than the $1.07\pm0.01$ ($1.11\pm0.03$) cross-calibration difference reported in \citet{madsen}. The ratio between the FPMA (FPMB) and {\it Chandra} normalizations for F2M 1227+3214 is $0.90^{+0.18}_{-0.15}$ ($0.81^{+0.17}_{-0.14}$), which is consistent within the uncertainties of the values reported between {\it NuSTAR} and {\it Chandra} grating spectroscopy in \citet{madsen}.\footnote{Cross-calibration between {\it NuSTAR} and ACIS CCD spectroscopy is not performed in \citet{madsen}.} The larger differences in the relative normalizations between detectors than those presented in \citet{madsen} in F2M 0830+3759 could be induced by uncertainties from the lower signal-to-noise in our spectra, compared with the bright calibration sources studied in \citet{madsen}. Indeed, past studies using joint spectral fitting of {\it NuSTAR} with {\it XMM-Newton} show similar cross-calibrational uncertainties to those we find here \cite[e.g.,][]{balokovic}. Additionally, AGN variability can play a role in the cross-calibrational differences.  We tested this by including the 9 ks {\it Chandra} spectrum in the joint fitting, finding the cross-calibration constant for {\it Chandra} relative to {\it XMM-Newton} (1.48+/-0.09) to be more consistent with that found between {\it NuSTAR} and {\it XMM-Newton}, suggesting that the source varies over time and was fainter during the epoch of the {\it XMM-Newton} observation, perhaps due to the direct emission that is scattered into our line-of-sight (see below for details). Errors on the spectral fit parameters are quoted at the 90\% confidence interval, corresponding to a $\Delta \chi^2$ of 2.7 for one interesting parameter.

\subsection{F2M 0830+3759}
We initially fit the {\it NuSTAR} and {\it XMM-Newton} spectra with an absorbed power law model, with an absorption component fixed to the Galactic value \citep[$N_{\rm H,Gal}$=$4\times10^{20}$ cm$^{-2}$;][]{gal_nh} and an additional component at the redshift of the quasar which was left free ($N_{\rm H,Z}$) :

\begin{equation}\label{pow_model}
\begin{split}
{\rm model = const} \times {\rm exp}[-N_{\rm H,Gal} \sigma(E)] \times \\
{\rm exp}[-N_{\rm H,Z}\sigma(E)]  \times A\times E^{-\Gamma},
\end{split}
\end{equation}

\noindent where $A$ is the normalization of the power-law, $\sigma(E)$ is the photoelectric cross-section, and the constant factor accounts for cross-calibration differences between {\it XMM-Newton} PN, MOS1 and MOS2 detectors and the {\it NuSTAR} FPMA and FPMB modules. Here, the absorption is modeled as a foreground screen of extinction. We note that though we write the equation in the rest-frame, we used XSpec model components {\it zphabs} and {\it zpowerlaw} here and below, where applicable, to appropriately account for redshift dependencies when modeling the observed-frame spectra and calculating flux.

This simple phenomenological model indicates at least moderate absorption ($N_{\rm H} > 10^{22}$ cm$^{-2}$). Additionally, 6.4 keV (rest-frame) Fe K$\alpha$ emission is also clearly evident, which is a signature of X-rays reflecting off either distant matter or the accretion disk \citep[e.g.,][]{krolik,george,shu,ricci}. We therefore fitted these spectra with physically motivated models that self-consistently account for the effects of photoelectric absorption, Compton scattering, and Fe K$\alpha$ fluorescence emission in the presence of moderate ($N_{\rm H} \sim 10^{22}$ cm$^{-2}$) to Compton-thick ($N_{\rm H} > 1.25\times10^{24}$ cm$^{-2}$) column densities. We model the intrinsic spectrum as a powerlaw, which is then modified by absorption and Compton scattering. Both the \citet{bn_torus} BNTorus and \citet{mytorus} MYTorus models realize these physical processes through a suite of Monte-Carlo simulations, producing pre-defined tables for input spectra with a range of physical parameters that are imported into XSpec for spectral fitting; such ``look-up'' tables allow the spectra to be fitted without having to integrate Monte Carlo results while modeling the spectra, which would result in impractical run times. MYTorus restricts the opening angle to 60$^{\circ}$ and fixes the Fe abundance to solar, while the BNTorus model allows the opening angle and Fe abundances to be free parameters; in both models, the inclination angle of the torus ($\theta_{\rm obs}$) can range from 0$^{\circ}$ (face-on) to 90$^{\circ}$ (edge-on), with the boundary between a face-on and edge-on geometry at 60$^{\circ}$ for MYTorus while this boundary between edge- and face-on depends on the torus opening angle in the BNTorus model. 

We note that \citet{liu2015} simulated X-ray torus reprocessing for Compton-thick column densities (10$^{24}$-10$^{25}$ cm$^{-2}$) for the geometries assumed by MYTorus and BNTorus in an attempt to reproduce the reported spectra of these models. They found that the latter model over-predicted the reflection component at low energies and over-predicted the Fe K$\alpha$ equivalent width for edge-on geometries relative to their calculations, while their results were fully consistent with the MYTorus model. However, \citet{brightman2015} simulated spectra in the energy range 3-79 keV with the BNTorus model for various torus opening angles and values of $\Gamma$. As they find that the MYTorus model fit to these simulated spectra recovers the input parameters for the case where the BNTorus model opening angle is 60$^{\circ}$, we include this model in the analysis below for completeness, freezing the opening angle to 60$^{\circ}$.

Soft excess emission is present in F2M 0830+3759 below 1 keV, which we attribute to AGN emission that ``leaks'' through the obscuring medium, either through the opening of the torus or holes in a clumpy obscuring medium, and either directly enters our line-of-sight or is subsequently scattered off a distant optically thin medium before transversing our line-of-sight. If such emission resulted from photons scattered by the torus itself, we would see signatures of this process in the reflected or transmitted spectrum, but modeling the spectra with just these processes fails to fit the soft emission. We therefore include a scattered powerlaw to our model to account for this component, similar to partial covering models used in previous works with phenemenological modeling \citep[e.g.,][]{winter,turner,lamassa2009,lamassa2011,mayo}.

The BNTorus model can be represented as:
\begin{equation}
\begin{split}
{\rm model = const} \times {\rm exp}[-N_{\rm H,Gal} \sigma(E)] \times \\
[{\rm BNTorus}(N_{\rm H},\Gamma,\theta_{\rm tor},\theta_{\rm obs},E) + \\
f_{\rm scatt} \times (A\times E^{-\Gamma})],
\end{split}
\end{equation}
\noindent where the BNTorus component depends on the equatorial column density ($N_{\rm H}$), the opening angle of the torus ($\theta_{\rm tor}$), the inclination angle of the torus ($\theta_{\rm obs}$), and energy ($E$) since the probability that a photon will undergo Compton-scattering depends on its incident energy. To preserve the self-consistency of the model, the powerlaw slope ($\Gamma$) and normalization ($A$) of the scattered emission are tied to the BNTorus values, with a constant multiplicative factor left free to measure the scattering fraction. The absorption due to the Galaxy \citep[$N_{\rm H,Gal}$=$4\times10^{20}$ cm$^{-2}$;][]{gal_nh}  is kept frozen. The first contant factor accounts for cross-calibration differences among the {\it XMM-Newton} PN, MOS1, and MOS2 detectors and the {\it NuSTAR} FPMA and FPMB detectors.

The MYTorus model instead has separate components for the transmitted, Compton-scattered, and fluorescent line emission:
\begin{equation}\label{myt_eq}
\begin{split}
{\rm model = const} \times {\rm exp}[-N_{\rm H,Gal} \sigma(E)] \times \\
[A\times E^{-\Gamma} \times {\rm MYTorusZ}(N_{\rm H,Z},\theta_{\rm obs},E)\ + \\
A_{\rm S} \times {\rm MYTorusS}(A,\Gamma,N_{\rm H,S},\theta_{\rm obs},E) + \\
A_{\rm L} \times {\rm MYTorusL}(A,\Gamma,N_{\rm H,S},\theta_{\rm obs},E) + \\
f_{\rm scatt} \times (A\times E^{-\Gamma})],
\end{split}
\end{equation}
where 
\begin{itemize}
\item MYTorusZ is the component that modifies the transmitted spectrum, where we used the MYTorus multiplicative table {\it mytorus\_Ezero\_v00.fits},
\item MYTorusS describes the Compton-scattered emission, where we have used the Monte-Carlo realization with a powerlaw termination energy of 200 keV (i.e., MYTorus table \\
{\it mytorus\_scatteredH200\_v00.fits})\footnote{As we work within an energy range far below the cut-off energy, the choice of MYTorus termination energy, which can range from 100-500 keV, has a small impact on our results} and
\item MYTorusL ({\it mytl\_V000010nEp000H200\_v00.fits}) accounts for fluorescent line emission.
\end{itemize}
All of these MYTorus components have a dependence on column density, inclination angle, energy, and redshift. To preserve the physical self-consistency, the powerlaw normalization and slope are tied together among the MYTorus components and scattering model during the fitting; $\theta_{\rm obs}$ and the column densities ($N_{\rm H,Z}$ and $N_{\rm H,S}$) are also tied among the MYTorus components. These constraints are required by the definition of the MYTorus model and input tables. The relative normalizations between the Compton-scattered emission ($A_{\rm S}$) and the fluorescent line emission ($A_{\rm L}$) are tied to each other and $A_{\rm S}$ is allowed to be free \citep[see e.g.,][]{mytorus,yaqoob2012,lamassa14}.

\begin{figure}[ht]
\begin{centering}
\includegraphics[scale=0.4]{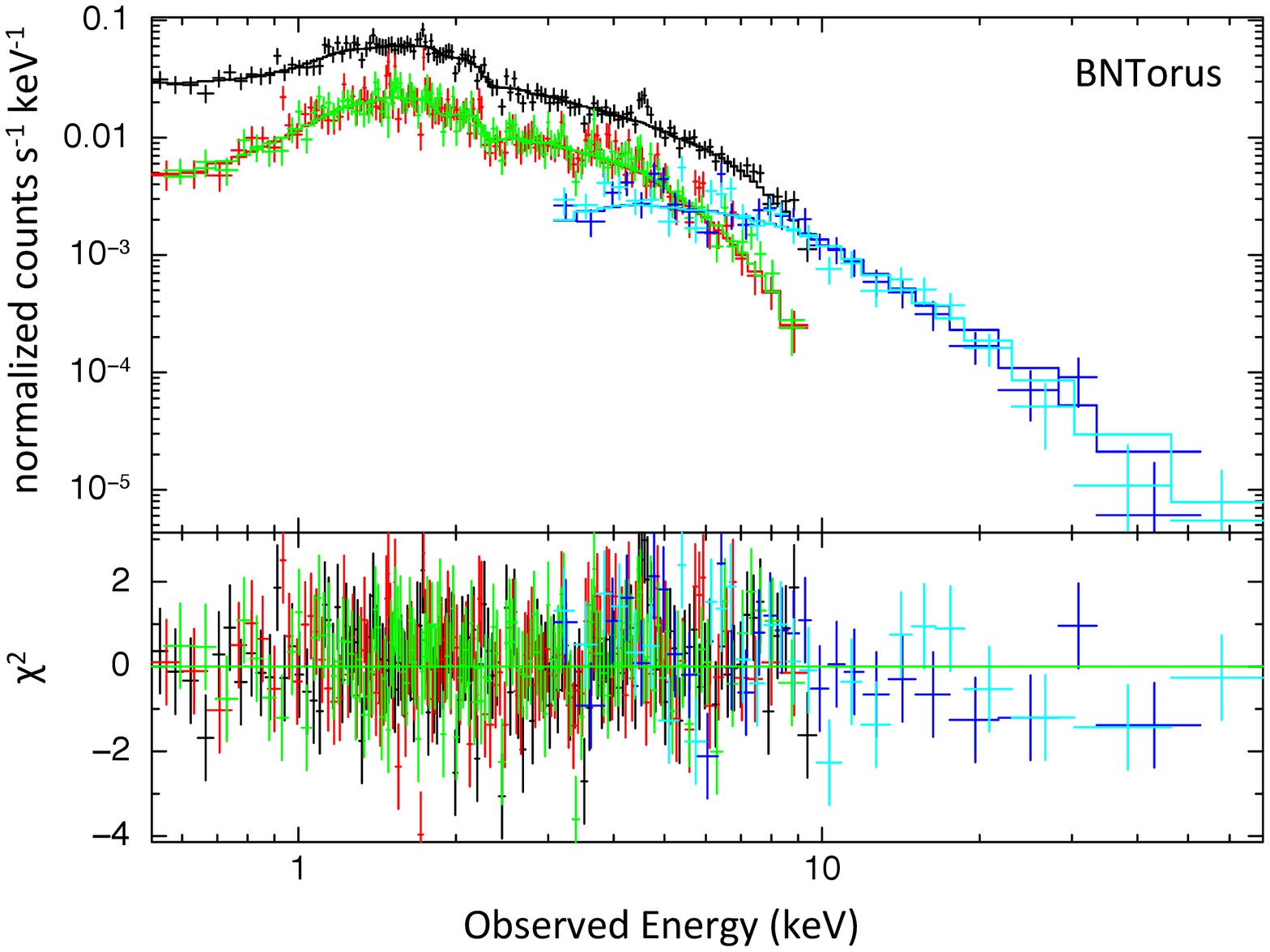}~
\hspace{1cm}
\vspace{1.1cm}
\includegraphics[scale=0.4]{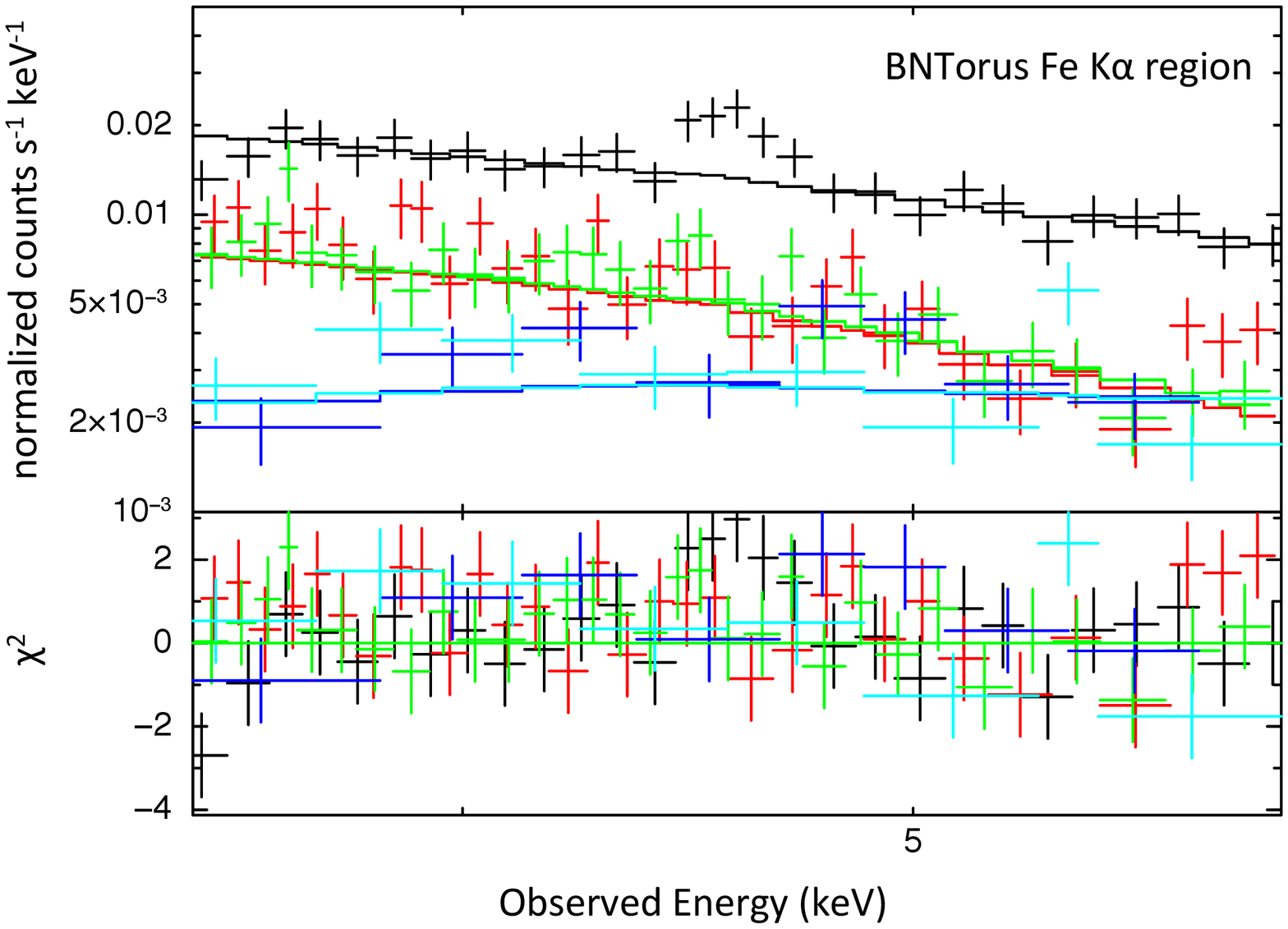}\\
\includegraphics[scale=0.4]{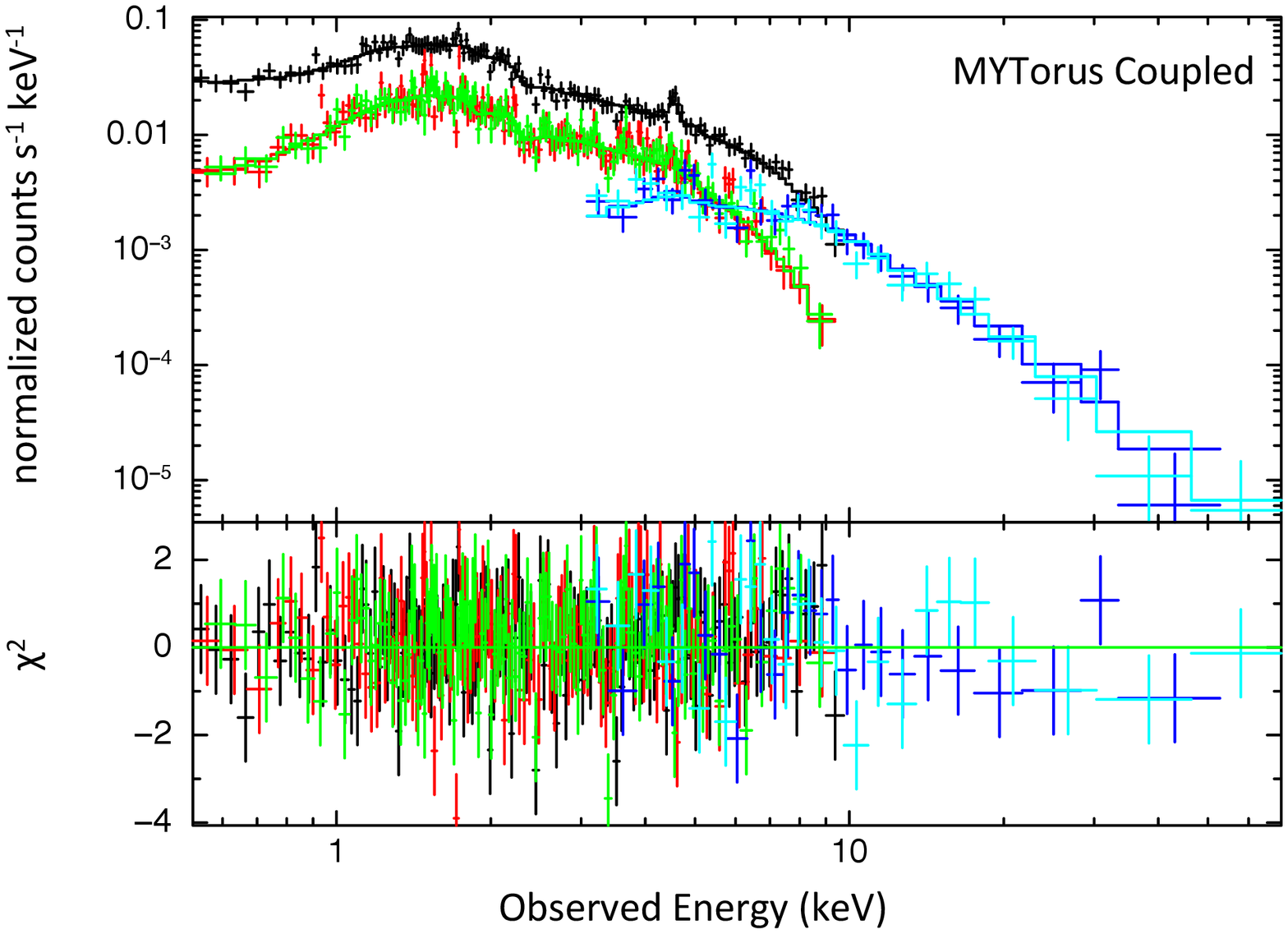}~
\hspace{1cm}
\includegraphics[scale=0.4]{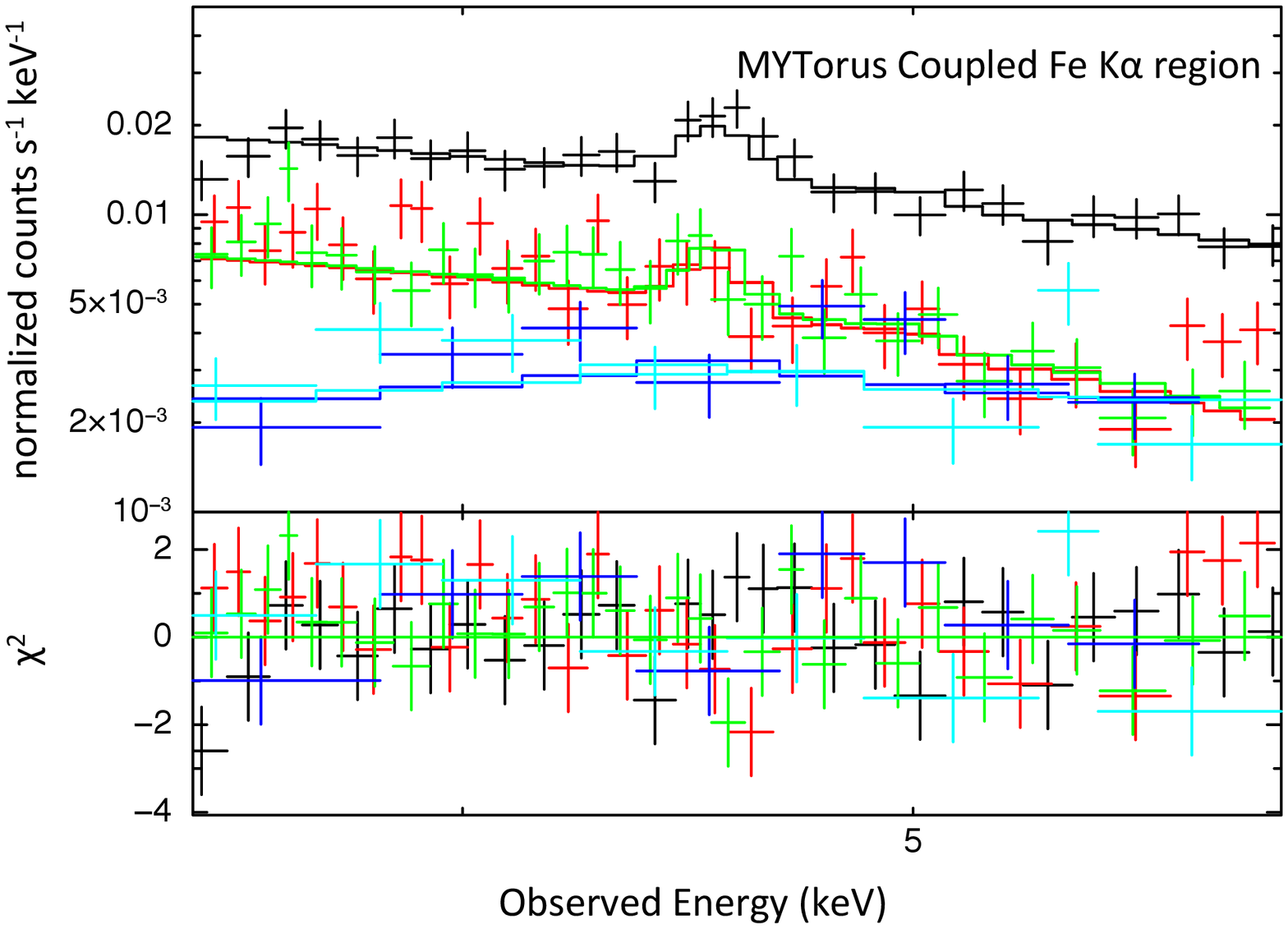}
\caption{\label{f2m0830_spec_coup} {\it Left}: X-ray spectra of F2M 0830+3759 ({\it XMM-Newton} PN  - black,  {\it XMM-Newton} MOS1 - red,  {\it XMM-Newton} MOS2 - green, {\it NuSTAR} FPMA - dark blue, FPMB - light blue) with the best-fit ``coupled'' toroidal models overplotted; $\chi^2$ residuals are plotted in the bottom panels. The BNTorus model fit is on top ($\chi^2$=473.4 for 433 degrees of freedom) while the MYTorus model is shown in the bottom row ($\chi^2$=444.4 for 432 degrees of freedom). {\it Right}: Close-up of the Fe K$\alpha$ region (rest-frame 6.4 keV) for both models. While the BNTorus model provides a good global fit, the Fe K$\alpha$ emission is poorly accommodated ({\it top}) compared with the MYTorus model ({\it bottom}). However, the fitted inclination angles of the torus suggest that the X-ray reprocessor is a more complex medium than described by these models which presume a homogeneous distribution of matter.}
\end{centering}
 \end{figure}

Both models provide a good global fit to the spectra (Figure \ref{f2m0830_spec_coup}, left), though the MYTorus model does a much better job of fitting the Fe K$\alpha$ complex than the BNTorus model (Figure \ref{f2m0830_spec_coup}, right). However, the fitted inclination angles suggest that complexities exist in the X-ray reprocessor that are not accounted for in the presumed geometry of the models. While $\theta_{\rm obs}$ is largely unconstrained in the BNTorus model fit (i.e., $>$62.5$^{\circ}$), due to the column density being independent of inclination angle in this model, it has a very narrow allowed range in the MYTorus fit (60.1$^{\circ} < \theta_{\rm obs} <$ 60.8$^{\circ}$), indicating a grazing incidence angle between the AGN continuum and the obscuring medium. This latter result indicates that the model is attempting to reconcile the competing effects of a strong reflection component, producing the Fe K$\alpha$ line, and a weakly absorbed transmitted continuum. These effects could result from multiple absorption components at physically different locations, suggesting that the X-ray reprocessor might not be a homogenous medium, or that a gradient in column density exists over the X-ray reprocessor such that the integrated emission provides a significantly different column density than that viewed along the line-of-sight. 

\begin{deluxetable}{llllllll}

\tablewidth{0pt}
\tablecaption{\label{fit_summary} {\it NuSTAR} \& {\it Chandra}/{\it XMM-Newton} Decoupled MYTorus Fit Summary}
\tablehead{\colhead{Source} & \colhead{$\Gamma$} & \colhead{$N_{\rm H,z}$\tablenotemark{1}} & \colhead{$A_{\rm S}$} & \colhead{$N_{\rm H,S}$} & \colhead{$f_{\rm scatt}$} & \colhead{$\chi^2$ (DOF)} \\
 & & \colhead{(10$^{22}$ cm$^{-2}$)} & & \colhead{(10$^{24}$ cm$^{-2}$)} }

\startdata
F2M 0830+3759 & 1.66$^{+0.07}_{-0.06}$ & 2.1$\pm0.2$ & 2.1$^{+3.2}_{-0.8}$ & 0.37$^{+0.41}_{-0.26}$ & 0.07$^{+0.09}_{-0.06}$ & 445.6 (432) \\
F2M 1227+3214 & 1.99$^{+0.12}_{-0.11}$ & 0.34$^{+0.08}_{-0.07}$ & 1\tablenotemark{2} & $<$0.55 & \nodata & 92.3 (95) \\
\enddata
\tablenotetext{1}{$N_{\rm H,z}$ represents the line-of-sight obscuration, while $N_{\rm H,S}$ denotes the global column density.}
\tablenotetext{2}{The $A_{\rm S}$ normalization was frozen to unity during the fitting since it was unconstrained when left free.}
\end{deluxetable}

We therefore fitted these spectra with MYTorus in ``decoupled'' mode, where the line-of-sight and global column densities are allowed to be distinct from each other and fit independently, consistent with a patchy obscuring medium. Here, a portion of the observed X-ray emission results from X-ray reflection off of the far-side of the obscuring medium that enters the line-of-sight without further interaction with the absorber. In this case, the global column density ($N_{\rm H,S}$) associated with this far-side reflection has an inclination angle frozen to 0$^{\circ}$ since it emulates the physics of a face-on torus. Conversely,  $\theta_{\rm obs}$ is fixed at 90$^{\circ}$ for the line-of-sight column density ($N_{\rm H,Z}$) since this describes the absorption that reprocesses the transmitted component. Unlike the coupled mode, the column densities are fit independently, where $N_{\rm H,S}$ is tied together between the MYTorusS and MYTorusL components. Again, the powerlaw parameters and relative normalizations ($A_{\rm S}$ and $A_{\rm L}$) are tied together and $A_{\rm S}$ is allowed to be free. We note that the fixed covering factor assumed in the MYTorus model does not greatly impact the derived column density values for covering factors below $\sim$0.8: the reflection spectrum for a face-on torus (i.e., the component associated with a 0$^{\circ}$ inclination angle) remains constant until the inclination angle intercepts the edge of the torus, or when the opening angle of the torus becomes narrow. As the covering factor approaches unity, the spectrum would best be described by, e.g., the spherical absorption model of \citet{bn_torus}. 

\begin{figure}[ht]
\begin{centering}
\includegraphics[scale=0.4]{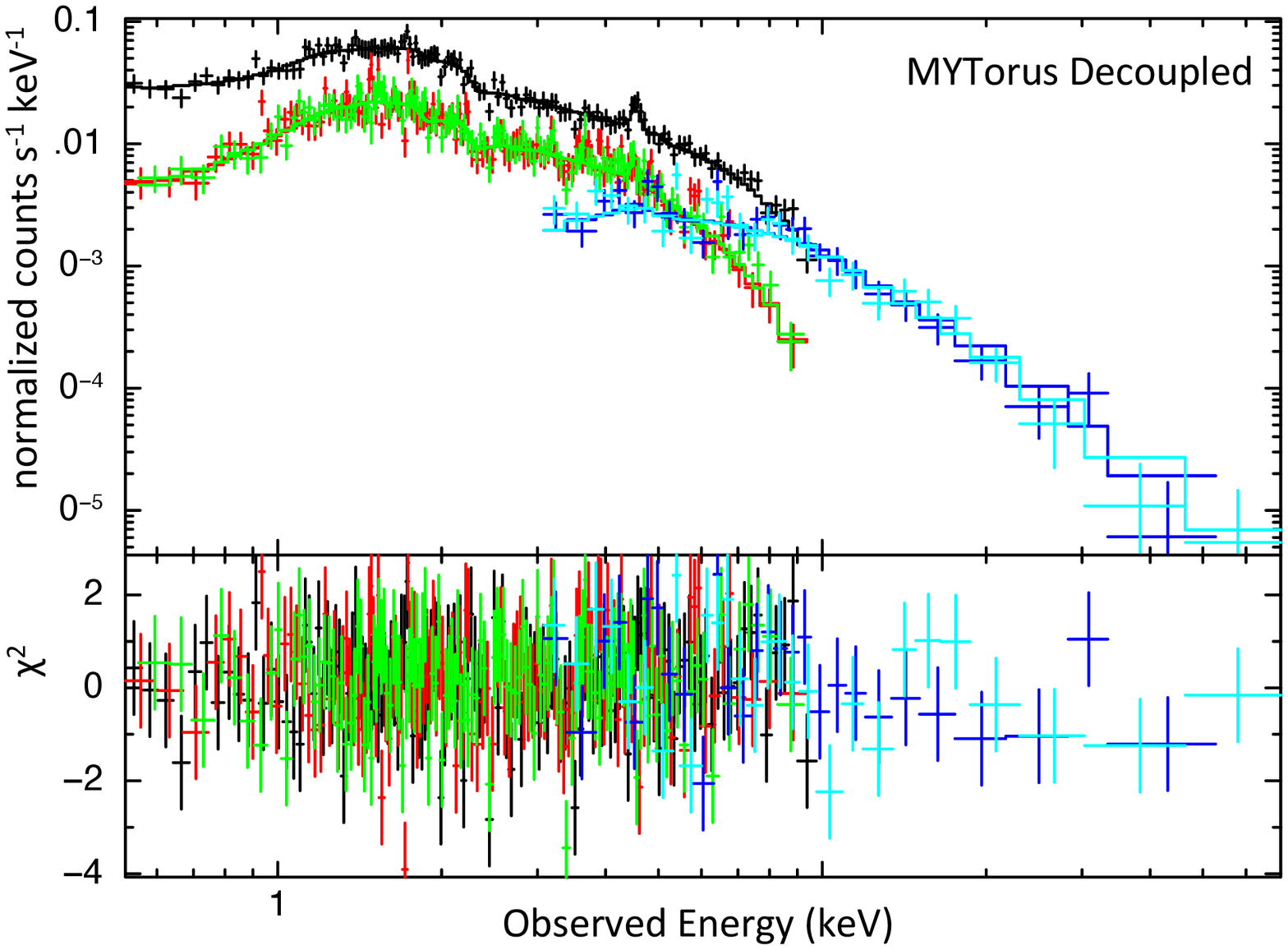}~
\hspace{1cm}
\includegraphics[scale=0.4]{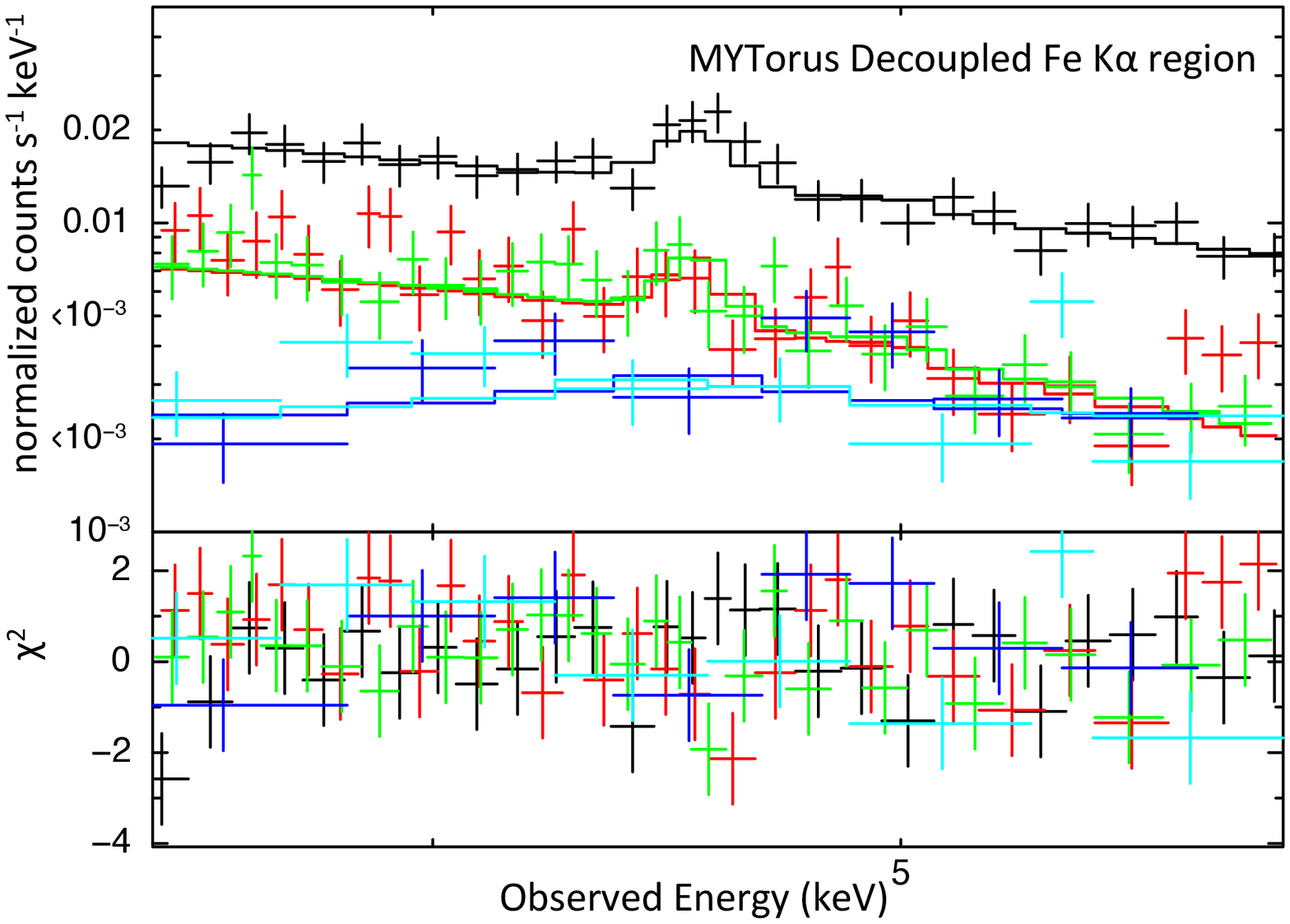}
\caption{\label{f2m0830_spec} X-ray spectra of F2M 0830+3759 (color coding same as Figure \ref{f2m0830_spec_coup}) with the decoupled MYTorus fit overplotted, with a close-up of the Fe K$\alpha$ region shown in the right-hand panel. In this realization of the MYTorus model, the line-of-sight and global column densities are disentangled from each other and fit independently, emulating a non-uniform and patchy obscuring medium.}
\end{centering}
 \end{figure}

\begin{figure}[ht]
\begin{centering}
\includegraphics[scale=0.45]{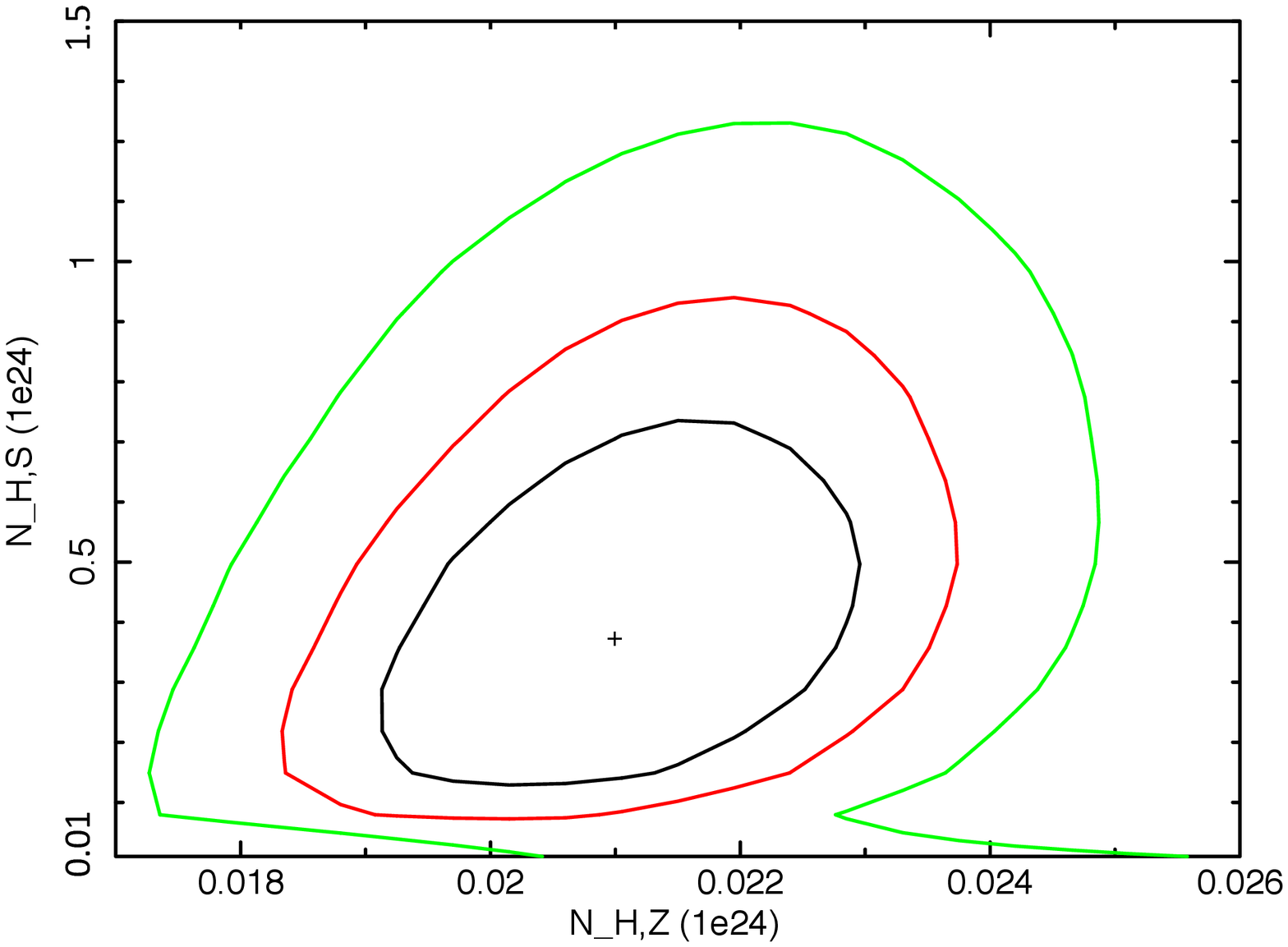}
\caption{\label{f2m0830_con} $\chi^2$ contour plots of the line-of-sight column density ($N_{\rm H,Z}$) versus the global column density ($N_{\rm H,S}$) for F2M 0830+3759, where the black, red, and green curves show the 68\%, 90\%, and 99\% confidence intervals respectively. While $N_{\rm H,Z}$ is constrained to be moderate ($1.7\times10^{22} < N_{\rm H,Z} < 2.5 \times10^{22}$ cm$^{-2}$), the global obscuration is much heavier ($N_{\rm H,S} > 10^{23}$ cm$^{-2}$) at the 90\% confidence interval.}
\end{centering}
 \end{figure}

The decoupled MYTorus fit to the observed spectrum, with a close-up of the Fe K$\alpha$ region, is shown in Figure \ref{f2m0830_spec}. As summarized in Table \ref{fit_summary}, the line-of-sight absorption is moderate ($N_{\rm H,Z} = 2.1\pm0.2 \times 10^{22}$ cm$^{-2}$) while the global column density is much higher, though not Compton-thick within the 90\% confidence level ($N_{\rm H,S} = 3.7^{+4.1}_{-2.6} \times 10^{23}$ cm$^{-2}$). We illustrate this further in Figure \ref{f2m0830_con}, where we show contour plots between the line-of-sight and global column densities. At the 68\% and 90\% confidence level, the colum densities are different, though future observations would be needed to improve the significance to the 99\% confidence interval. We found that the higher energy coverage of {\it NuSTAR} plays a critical role in determining the global column density as the upper limit on $N_{\rm H,S}$ is unconstrained with only the {\it XMM-Newton} data, as we discuss in more detail below. We note that the $N_{\rm H,S}$ global column density does not necessarily mean that this obscuration is on galactic-sized scales. Rather, this absorption represents gas near the black hole that plays a role in reprocessing the AGN emission, but does not intersect the direct view to the central engine. 

Finally, we find that $\sim$7\% of the intrinsic AGN continuum ``leaks'' through the patchy obscuring medium and is subsequently scattered into our line-of-sight. When modeling this soft excess emission with a thermal component (the {\it apec} model in XSpec), which would be appropriate if this emission is due to star-formation, instead of a scattered powerlaw model, we find that in order to find a good fit to the data ($\chi^2$=472.2 for 431 DOF), $A_{\rm S}$ becomes largely unconstrained, with an unphysical nominal value ($5.6^{+8.5}_{-3.6}$), suggesting that scattered AGN light is the more likely source of this emission.

As noted above, there could be an enhancement of X-ray emission due to putative jet-linked radiation which could contribute to the light which we have interpreted as being due to leakage of the intrinsic AGN continuum through the circumnuclear medium or could dilute the Fe K$\alpha$ line and reflection component, thereby affecting the line-of-sight column measurement. Our data, however, are not of high enough quality to determine whether this possible contamination exists and could be disentangled from the remaining X-ray emission.

\subsubsection{Comparison with Previous X-ray Analysis}
F2M 0830+3759 was first observed in X-rays with {\it Chandra} with an exposure time of $\sim$9 ks \citep{urrutia}. The Fe K$\alpha$ line was detected in this observation and was assumed to be a Doppler broadened line, where they find a line width of $0.6\pm0.3$ keV.  They also obtain a much steeper photon index ($\Gamma=2.9 \pm 0.1$) than we find here and steeper than found by \citet{piconcelli} in their re-analysis of the F2M 0830+3759 {\it Chandra} spectrum ($\Gamma=1.65\pm0.25$).

\citet{piconcelli} reported the $\sim$50 ks {\it XMM-Newton} observation of F2M 0830+3759, which resulted in a much higher quality spectrum than previously available. They fitted the soft-excess below 1 keV in a similar way as reported here, albeit with phenomenological modeling instead of the physically motivated BNTorus \citep{bn_torus} and MYTorus \citep{mytorus} models since they were unavailable at the time. To accommodate the soft-excess, they included an additional power law component with the same photon index as the intrinsic continuum, finding a best-fit $\Gamma$ of 1.51$\pm$0.06 and $f_{\rm scatt}$ = 0.13$\pm$0.02. Though they find a good statistical fit to the spectrum with this model, they note that a marginally better fit is obtained when fitting a photoionized absorber model with XSTAR\footnote{In this model, the absorption is dependent on the ionization parameter, which is defined by $L/nr^2$, where $L$ is the luminosity of the ionizing source, $n$ is the plasma density, and $r$ is the radial distance between the source and the absorber.} \citep{xstar}. In both cases, this absorption is along the line-of-sight and the soft-excess likely results from distant scattering of the intrinsic AGN continuum that leaks through openings in the absorbing medium. \citet{piconcelli} do not interpret the Fe K$\alpha$ line to be Doppler broadened. They calculate a rest-frame Fe K$\alpha$ EW of 168$\pm$60 eV, which is similar to the value we obtain ($\sim$169 eV). Here, we are able to extend the \citet{piconcelli} analysis further, inferring both the line-of-sight and global column densities, and relaxing the inherent assumption of infinite column density in the {\it pexrav} reflection model they used.

\subsection{F2M 1227+3214}

\begin{figure}[ht]
\begin{centering}
\includegraphics[scale=0.45]{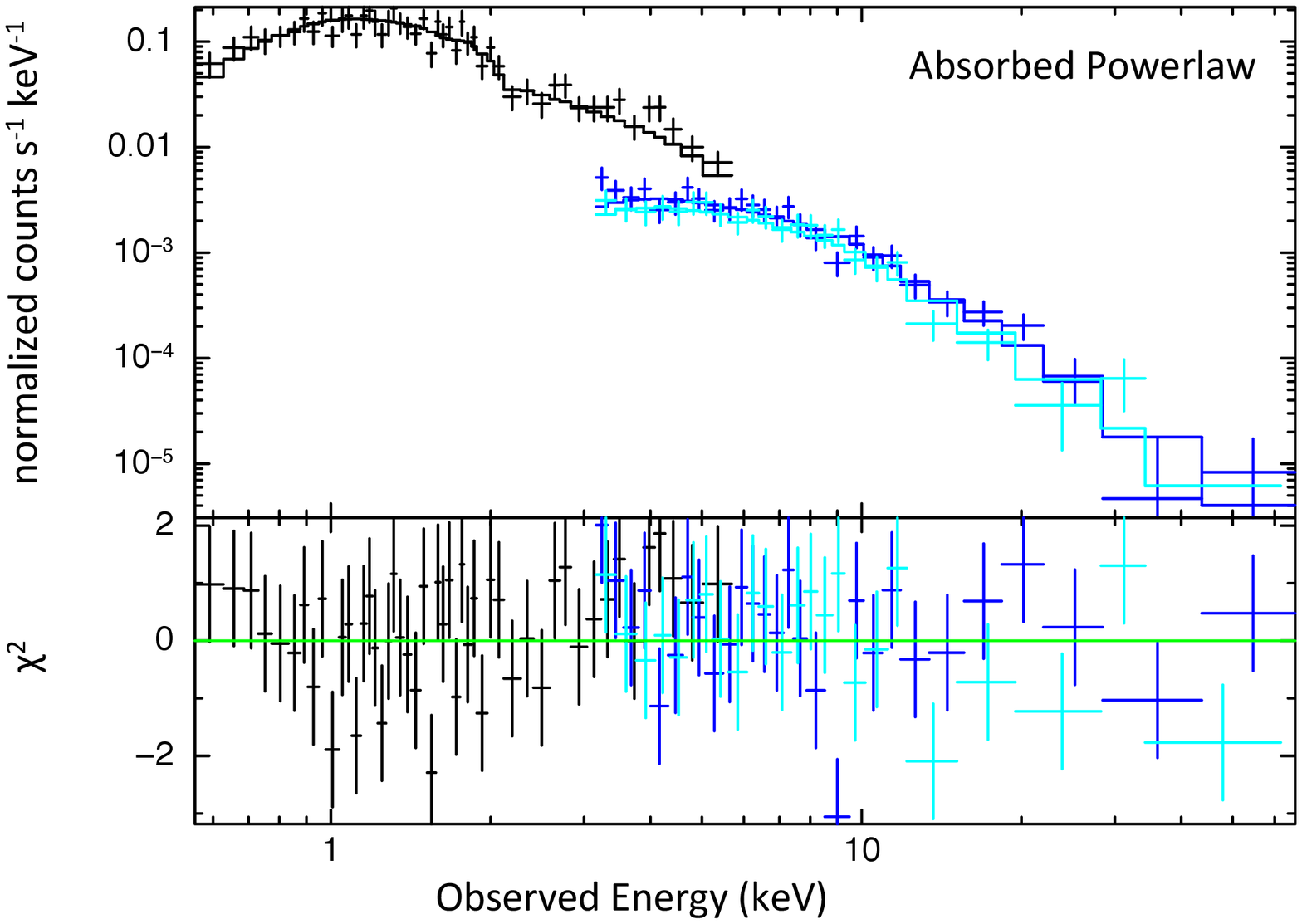}
\caption{\label{f2m1227_spec} Spectra and absorbed power law model fit for F2M 1227+3214 with $\chi^2$ residuals in the bottom panel ($\chi^2$=92.8 for 96 degrees of freedom); the {\it Chandra} spectrum is black and the {\it NuSTAR} spectra are red (FPMA) and green (FPMB). This spectrum is typical for a mildly absorbed Type 1 AGN. However, when fitting the spectrum with MYTorus in decoupled mode, which fits the data equally well ($\chi^2$=92.3 for 95 degrees of freedom), we find an upper limit on the global column density (out of the line-of-sight) of $5.5\times10^{23}$ cm$^{-2}$.}
\end{centering}
 \end{figure}

We fitted the {\it Chandra} and {\it NuSTAR} data for this source using an absorbed power-law model as parameterized in Equation \ref{pow_model}. Here, the Galactic absorption is $N_{\rm H,Gal} = 1.7\times10^{20}$ cm$^{-2}$ \citep{gal_nh}. As shown in Figure \ref{f2m1227_spec}, this simple model provides a good fit to the data, and is typical of a Type 1 AGN spectrum, with $\Gamma$=1.99$^{+0.12}_{-0.11}$ \citep[e.g.,][]{manieri,lanzuisi} and a mild absorption along the line-of-sight of $N_{\rm H,Z}$=3.4 $^{+0.8}_{-0.7}\times10^{21}$ cm$^{-2}$.

To test whether there may be higher global obscuration out of the line-of-sight, we fit the spectra of F2M 1227+3214 with MYTorus in decoupled mode. Here, we replaced the MYTorusZ component, which modifies the zeroth-order transmitted continuum, with {\it zphabs} $\times$ {\it zpowerlaw} since the line-of-sight column density is lower than that capable of being modeled with MYTorus ($N_{\rm H,Z, min} = 10^{22}$ cm$^{-2}$). This replacement is justified since Compton-scattering has a negligible impact on the shape of the transmitted spectrum for line-of-sight column densities under 10$^{22}$ cm$^{-2}$. Fitting the spectra with MYTorusZ causes the model to underpredict the observed emission at energies $<$2 keV due the minimum attenuation for the model being too high compared with the observed spectrum. Similar to the decoupled mode MYTorus set-up, the inclination angle of the MYTorusS and MYTorusL components is frozen at 0$^{\circ}$. From this fitting, we derive a 90\% confidence level upper limit on the global column density of $<5.5\times10^{23}$ cm$^{-2}$. The fit parameters from this modeling are summarized in Table \ref{fit_summary}.

\FloatBarrier

\begin{deluxetable}{llllll}
\tablewidth{0pt}
\tablecaption{\label{xray_flux}Net Counts\tablenotemark{1} and Observed X-ray Fluxes\tablenotemark{2} (10$^{-12}$ erg cm$^{-2}$ s$^{-1}$)}
\tablehead{\colhead{Source} & \colhead{Counts} & \colhead{Counts} & \colhead{$F_{\rm 2-10keV}$} & \colhead{$F_{\rm 10-40keV}$} & \colhead{$F_{\rm 2-40keV}$} \\
 & \colhead{{\it Chandra}/{\it XMM-Newton}} & \colhead{{\it NuSTAR} }}

\startdata
F2M 0830+3759 & 12433$\pm$112 & 1169$\pm$34 & 0.95$^{+0.13}_{-0.12}$ & 1.78$^{+0.25}_{-0.23}$ & 2.90$^{+0.40}_{-0.38}$ \\
F2M 1227+3214 & 834$\pm$29       & 1082$\pm$33 &1.30$^{+0.19}_{-0.17}$ & 1.10$^{+0.16}_{-0.14}$ & 2.33$^{+0.35}_{-0.30}$ \\
\enddata
\tablenotetext{1}{The net counts for F2M 0830+3759 as detected by {\it XMM-Newton} corresponds to the energy range 0.5-10keV, added among the PN, MOS1, and MOS2 detectors, and to the 0.5-8 keV band for F2M 1227+3214, as observed by {\it Chandra}. For both sources, the {\it NuSTAR} net counts are reported in the 3-79 keV range and are summed between the FPMA and FPMB detectors.}
\tablenotetext{2}{The errors refer to the statistical errors from the modeling, which are lower than the systematic errors on the absolute flux, which depend on the accuracy of the instruments and cross-calibration uncertainties.}
\end{deluxetable}

\begin{deluxetable}{llll}
\tablewidth{0pt}
\tablecaption{\label{xray_lum}Intrinsic, Rest-Frame X-ray Luminosities\tablenotemark{1} (erg s$^{-1}$)}
\tablehead{\colhead{Source} & \colhead{$L_{\rm 2-10keV}$} & \colhead{$L_{\rm 10-40keV}$} & \colhead{$L_{\rm 2-40keV}$} }

\startdata
F2M 0830+3759 & 44.84$\pm$0.06 & 45.11$\pm$0.06 & 45.35$\pm$0.06 \\
F2M 1227+3214 & 43.85$\pm$0.06 & 43.79$\pm$0.06 & 44.12$\pm$0.06 \\

\enddata
\tablenotetext{1}{The luminosities are reported in log space.}
\end{deluxetable}

\subsubsection{Effects of Variability}
As the {\it NuSTAR} observations are non-contemporaneous with the archival {\it XMM-Newton} and {\it Chandra} observations, variability could affect our spectral modeling and the derived column density and absorption-corrected luminosity values. Here we explore the extent of such possible effects by fitting the spectra from the lower energy and higher energy observations independently to determine whether the results are inconsistent with those found from joint fitting.

When we apply the coupled MYTorus model to the {\it XMM-Newton} spectra of F2M 0830+3759, we find similar results from the joint fitting, where the inclination of the reprocessor is constrained to be at a grazing incidence angle. Conversely, the fit to the {\it NuSTAR}-only spectra results in an inclination angle that is completely unconstrained. We then fit these spectra independently with the MYTorus model in decoupled mode and find that the global column density is unconstrained in both cases. The measured line-of-sight column density is 2.0$\pm0.2\times10^{22}$ cm$^{-2}$ when modeling the  {\it XMM-Newton} only spectra while it has a much larger allowed range when fitting just the {\it NuSTAR} spectra ($N_{\rm H,Z}=14^{+15}_{-12}\times10^{22}$ cm$^{-2}$). Though there is a wide range of allowed column density values, the results from this independent modeling do not contradict the values we derive when fitting the spectra jointly.

For F2M 1227+3214, we applied the modified MYTorus decoupled model discussed above, where the MYTorusZ component was replaced with an absorbed powerlaw model to attenuate the transmitted emission, since the MYTorus models have a lower column density limit of 10$^{22}$ cm$^{-2}$ which is higher than the line-of-sight column density for this source. When fitting the {\it Chandra} and {\it NuSTAR} data independently, we find that the global column density is completely unconstrained. While we are able to measure the line-of-sight column density when fitting the {\it Chandra} spectrum, finding it in agreement with the value derived from the joint fit ($N_{\rm H,Z} = 0.19^{+0.09}_{-0.08}\times10^{22}$ cm$^{-2}$), it is consistent with zero when modeling just the {\it NuSTAR} data; this latter result is expected since the line-of-sight obscuration is too weak to impact the harder X-ray emission, so it has no measurable effect on the spectrum.

The results of these exercises indicate that any variability present between observational epochs is within the allowed parameter space that would be inferred from any given epoch. The broad-band 0.5-79 keV coverage allow us to place much tighter constraints on the physical properties of the X-ray reprocessor than we would be able to obtain with coverage in only one band. However, as the allowed values, especially for the global column densities, span a wide range, it is possible that the more precise values we obtain from the joint fitting do not reflect a constant column density between epochs since we are unable to rule out variation within the permitted ranges. 

Indeed, even contemporaneous lower and higher energy coverage could be limited in its utility to determine whether obscuration changes occurred. As pointed out by \citet{marinucci} who studied the {\it XMM-Newton} and {\it NuSTAR} spectra of nearby Compton-thick AGN NGC 1068, the $<$10 keV spectra can exhibit no changes over time, yet the spectrum above 10 keV can vary. During a monitoring campaign where NGC 1068 was observed jointly with {\it XMM-Newton} and {\it NuSTAR}, they found that the {\it XMM-Newton} spectra remained constant, but that the {\it NuSTAR} spectrum varied between 2012 \citep[reported in][]{bauer} and 2014, after which the source returned to the previous state in 2015. Assuming the circumnuclear obscuring medium takes the form of a patchy distribution, they attribute the excess emission above 10 keV observed in 2014 to be due to a cloud moving out of the line-of-sight, changing the effective column density by more than $2.5\times10^{24}$ cm$^{-2}$, unveiling the central engine at higher energies. As we do not have observations above 10 keV at earlier times for the FIRST-2MASS red quasars, we are unable to test for such an effect in our data. However, unlike NGC 1068, the normalization of the spectrum for F2M 0830+3759 at energies below 10 keV did vary between the {\it Chandra} and {\it XMM-Newton} observations, though the spectral shape, through which the column density is determined, stayed constant. In this case, even in the presence of flux variability, the measured column density affecting the lower energy spectrum shows no evidence of significantly changing.

\section{Discussion}
From the fits to the X-ray spectra we derive the observed X-ray fluxes (Table \ref{xray_flux}) and rest-frame intrinsic (i.e., absorption and reflection corrected) luminosities (Table \ref{xray_lum}). The reported errors reflect the statistical error of the fit (i.e., the uncertainty of ther powerlaw normalization at the 90\% confidence level), which is lower than the systematic errors due to absolute calibration of the detectors. Both objects have intrinsic X-ray luminosities consistent with quasars (i.e., $L_{\rm x} > 10^{44}$ erg s$^{-1}$). However, the X-ray spectral properties are quite different between the two sources. While F2M 1227+3214 has a simple X-ray spectrum well characterized by a single absorbed power law model with mild absorption, applying the MYTorus decoupled mode to this source indicates that the global column density may be up to two orders of magnitude higher. Though the spectra of F2M 1227+3214 has hints that the global column density may be much higher than that along the line-of-sight, the spectrum of F2M 0830+3759 requires that this be the case. 

Such significantly different column densities globally versus along the line-of-sight, where the former is much higher than the latter, is consistent with the expecations of the red quasar paradigm: these sources may be in the midst of expelling their cocoons of obscuring gas, making the view to the central engine relatively unobscured and broad-emission line gas visible, while large gas column densities out of the line-of-sight are still present and play a role in reprocessing the observed X-ray emission. Indeed, the fact that the global column density is shown by our X-ray observations to be below the Compton-thick regime is consistent with the picture presented in \citet{glikman2012}, where the red quasars are observed in a phase after the Compton-thick gas is evacuated, as in the \citet{hopkins2005} model. Additionally, if these systems were viewed from another angle, namely through the heavier global columns of gas, they may possibly be viewed as Type 2 (narrow-line) quasars, analagous to those discovered in SDSS \citep{zakamska,reyes,jia,lansbury1,lansbury2}. Since the soft excess is well accommodated by a model in which the AGN continuum is scattered, this suggests that this emission arises from physical processes associated with black hole fueling rather than other activity within the host galaxy.

Below, we relate the X-ray properties to the optical and infrared characteristics of these sources.

\subsection{Optical Reddening versus X-ray Obscuration}
\begin{figure}[ht]
\begin{centering}
\includegraphics[scale=0.4,angle=90]{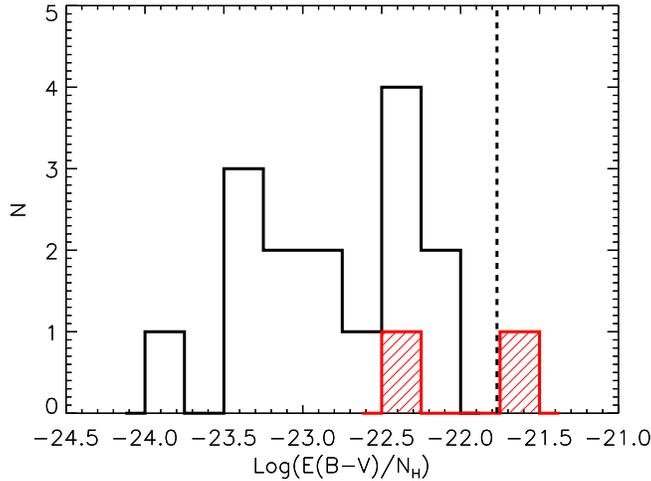}
\caption{\label{nh_ebv} Distribution of inferred dust-to-gas ratios (parameterized by $E(B-V)$/$N_{\rm H,Z}$) for the FIRST-2MASS quasars analyzed in this study (red) and the L$_{\rm 2-10keV}>10^{42}$ erg s$^{-1}$ sample presented in \citet{maiolino}, shown in black; the Galactic standard is shown by the dashed line. While F2M 1227+3214 has an $E(B-V)$/$N_{\rm H}$  value consistent with Galactic, F2M 0830+3759 is below this value, though on the high end of the sample considered in \citet{maiolino}.}
\end{centering}
 \end{figure}

Here, we compare the optical reddening in these quasars, as quantified by their $E(B-V)$ values derived above, with their X-ray obscuration determined by their fitted $N_{\rm H}$ values. \citet{maiolino} reported that the $E(B-V)$/$N_{\rm H}$ values for nearby AGN are significantly different from the Milky Way value \citep[$1.7\times10^{-22}$ mag cm$^2$;][]{bohlin}, and that there are systematic differences depending on intrinsic 2-10 keV luminosity: more luminous AGN with L$_{\rm 2-10keV}>10^{42}$ erg s$^{-1}$ have systematically lower $E(B-V)$/$N_{\rm H}$ values relative to the Galactic standard by factors of several to a hundred, while low-luminosity AGN ($L_{\rm 2-10keV} < 10^{42}$ erg s$^{-1}$) have higher values than observed in the Milky Way (albeit with only three objects in their sample in the latter group). In Figure \ref{nh_ebv}, we compare the $E(B-V)$/$N_{\rm H}$ values for the red quasars with the L$_{\rm 2-10keV}>10^{42}$ erg s$^{-1}$ sample from \citet{maiolino}. In that work, \citet{maiolino} compare $E(B-V)$ to the line-of-sight column density when the spectra require multiple absorption components to be fitted with a partial covering model. To be consistent with this practice, $E(B-V)$/$N_{\rm H}$ is calculated using the line-of-sight column density for the FIRST-2MASS quasars presented here.

F2M 1227+3214 has an $E(B-V)$/$N_{\rm H}$ value roughly consistent with the Galactic standard, while F2M 0830+3759 is lower, though on the higher end of the sample presented in \citet{maiolino}. For this latter source, the discrepancy between the measured and Galactic $E(B-V)$/$N_{\rm H}$ values could be due to physically disparate scales between the dust attenuating the optical emission and the gas obscuring and reprocessing the X-ray emission. Indeed, though we have a direct view of the broad line region in F2M 0830+3759, {\it Hubble} images show that it resides in the remnant of a merger \citep{urrutia2008}, where dust tends to be distributed on galactic, rather than circumnuclear, scales. Within the dust sublimation zone of the broad-line region, gas can attenuate and reprocess the X-ray emission \citep[e.g.,][]{risaliti2009,risaliti2010,risaliti2011,maiolino2010}.

\subsection{The $L_{\rm X}$-$L_{\rm 6\mu m}$ Plane}
\begin{figure}[ht]
\begin{centering}
\includegraphics[scale=0.35,angle=90]{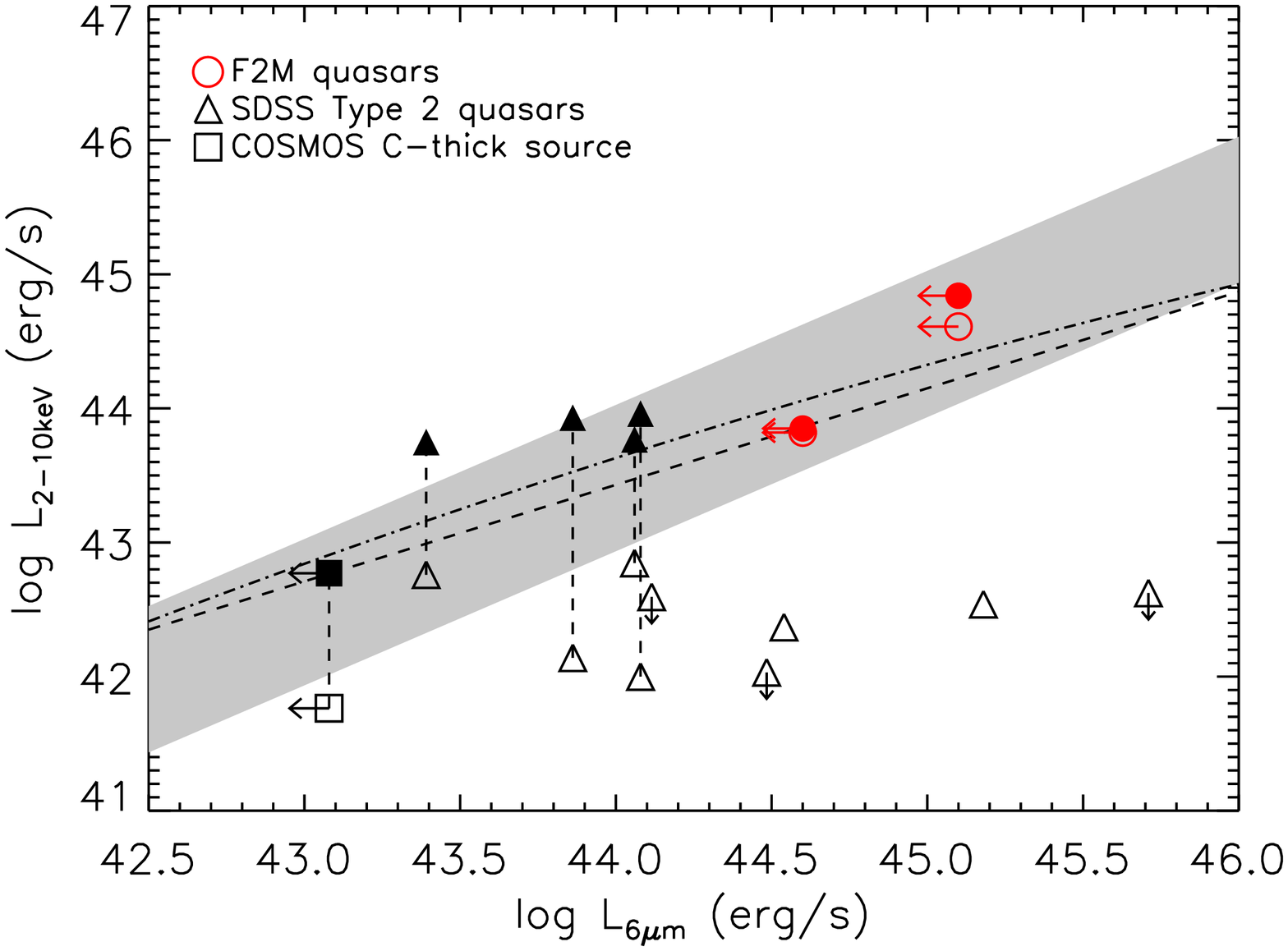}~
\includegraphics[scale=0.35,angle=90]{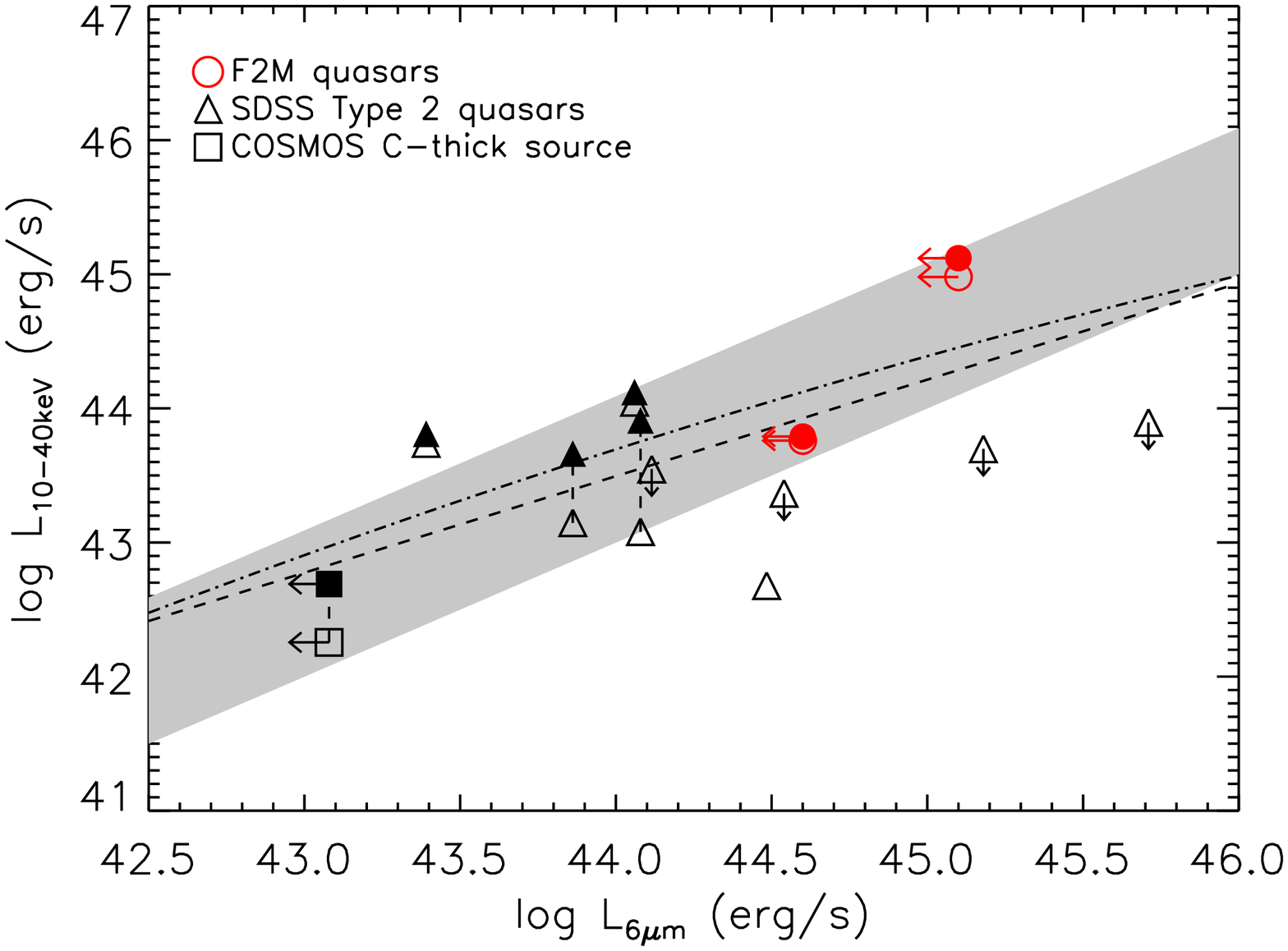}
\end{centering}
\caption{\label{mir_v_x} Rest-frame 6$\mu$m luminosity vs. rest-frame, non-absorption corrected (open symbols) and intrinsic (filled symbols) ({\it left}) 2-10 keV and ({\it right}) 10-40 keV luminosities for the FIRST-2MASS red quasars (red circles), SDSS Type 2 quasars (black diamonds) from \citet{lansbury1,lansbury2} and a Compton-thick source from the {\it NuSTAR} COSMOS survey, ID 330 \citep{civano}. The mid-infrared luminosities of the SDSS Type 2 quasars are estimates of the AGN heated dust from SED modeling while $L_{\rm 6\mu m}$ is the total 6$\mu$m emission for the red quasars and the Compton-thick source from COSMOS. The grey shaded region shows the empirical relationship between mid-infrared and X-ray luminosity for Seyfert galaxies from \citet{lutz} while the dashed-line and dot-dashed line shows a flatter relationship, which is derived from observations of luminous Type 1 quasars \citep[][respectively]{fiore2009,stern2015}. While the SDSS Type 2 quasars are observed to be underluminous in 2-10 keV X-rays compared with their mid-infrared emission, likely due to heavy levels of obscuration, the FIRST-2MASS red quasars are much more X-ray luminous since they have milder obscuration along the line-of-sight. The COSMOS Compton-thick AGN is X-ray under-luminous compared with the SDSS Type 2 and FIRST-2MASS quasars.}
 \end{figure}

A significant fraction of the mid-infrared emission in AGN arises from circumnuclear dust heated by the central engine, making such reprocessed emission a tracer of the intrinsic AGN power \citep[e.g.,][]{spinoglio,lamassa2010}. Relationships between the observed X-ray emission and the mid-infrared luminosity can then give a sense of the AGN obscuration \citep[e.g.,][]{alexander2008,gandhi,goulding2011,lamassa2009,lamassa2011,stern2014,lansbury1,lansbury2}. However, as pointed out by \citet{yaqoob2011} the ratio of the X-ray to mid-infrared luminosity is also strongly affected by the steepness of the AGN X-ray continuum and global covering fraction of the obscuring medium, so we caution that this ratio is not a clean diagnostic of X-ray obscuration. \citet{lutz} presented a relationship between absorption-corrected 2-10 keV luminosities and 6$\mu$m luminosities for Type 1 and Type 2 Seyfert galaxies (grey shaded region in Figure \ref{mir_v_x}, left). However, this relation appears to flatten at higher luminosities when samples of more distant and more luminous unobscured AGN from COSMOS \citep[dashed line in Figure \ref{mir_v_x};][]{fiore2009} and SDSS \citep[dot-dashed line in Figure \ref{mir_v_x};][]{stern2015} are considered. 

In Figure \ref{mir_v_x} we compare the rest-frame X-ray luminosities with the rest-frame mid-infrared 6$\mu$m luminosities of the FIRST-2MASS quasars, the SDSS Type 2 quasar candidates studied in \citet{lansbury1,lansbury2} and the Compton-thick source, ID 330, discovered in the {\it NuSTAR} COSMOS survey \citep{civano}.\footnote{The \citet{lutz}, \citet{fiore2009}, and \citet{stern2015} relations in the right-hand panel of Figure \ref{mir_v_x} are estimated by assuming an X-ray power-law continuum with $\Gamma$=1.8.} The 6$\mu$m luminosities for the SDSS Type 2 quasars represent emission from the AGN heated dust, estimated by fitting the SEDs of these sources \citep{lansbury1,lansbury2} while the 6$\mu$m luminosities for the Compton-thick COSMOS AGN and the FIRST-2MASS quasars are the total mid-infrared emission, including that from the host galaxy.\footnote{Such host galaxy mid-infrared emission is likely negligible for the FIRST-2MASS sources \citep[c.f.,][]{stern2015}.} The open symbols represent the non-absorption corrected X-ray luminosities while the filled symbols are intrinsic X-ray luminosities, if data exist to calculate this quantity; dashed lines connect the absorbed and intrinsic X-ray luminosities for the same source. We note, however, that the absorption corrected X-ray luminosity is based on the column densities measured from the joint fitting of spectra obtained during different epochs, and the column density could have varied between the epochs and/or between the infrared and X-ray observations.

While the SDSS Type 2 quasars and the Compton-thick COSMOS AGN tend to have absorbed 2-10 keV X-ray luminosities below that predicted by their mid-infrared luminosities, several of these objects have 10-40 keV emission more consistent with the empirical relations derived by \citet{lutz}, \citet{fiore2009}, and \citet{stern2015}. This result is consistent with the paradigm that obscuration is suppressing the lower energy X-ray emission while higher energy X-rays pierce through the high columns of gas, suffering much less attenuation. Indeed, where data are available to correct the X-ray luminosities for absorption, the SDSS Type 2 quasars have intrinsic X-ray luminosities similar to the FIRST-2MASS quasars. Both of the red quasars, however, have 2-10 keV luminosities consistent with the $L_{\rm 6\mu m}$-$L_{\rm 2-10keV}$ relations derived for luminous quasars by \citet{fiore2009} and \citet{stern2015}. Unlike the SDSS Type 2 quasars and the Compton-thick source from COSMOS, the difference between the intrinsic and absorbed X-ray luminosities is not extreme since the line-of-sight obscuration is mild to moderate.

\section{Conclusions} 
We have presented the X-ray analysis, including {\it NuSTAR} data, of two red quasars, F2M 0830+3759 and F2M 1227+3214, representing the first detection of theses sources at energies above 10 keV. Such red quasars, selected from the FIRST and 2MASS surveys, are hypothesized to be a transitional link between heavily enshrouded supermassive black hole growth caused by major galaxy mergers and the traditional Type 1 quasars efficiently discovered in optical surveys \citep{glikman2007,glikman2012,glikman2013,brusa2005,brusa2007,banerji2012,banerji2015}. Indeed, {\it Hubble} imaging of F2M 0830+3759 reveals a train-wreck host galaxy, evidence of a past major merger \citep{urrutia2008}. Additionally, after correcting for reddening, these quasars are among the most luminous AGN at every redshift \citep{glikman2012,banerji2015}, though less extreme than the {\it WISE}-discovered Hot DOGs \citep{assef2015,tsai}.  We summarize the main results below, where both the broad-band X-ray coverage (0.5 - 79 keV) from {\it Chandra}/{\it XMM-Newton} and {\it NuSTAR} and physically motivated X-ray models are crucial for providing clear insight into the physical processes at play in these luminous obscured AGN:

\begin{itemize}
\item Both F2M 0830+3759 and F2M 1227+3214 have mild-to-moderate absorption along the line-of-sight ($N_{\rm H,Z}$=$2.1\pm0.2\times10^{22}$ cm$^{-2}$ and $3.4^{+0.08}_{-0.07}\times10^{21}$ cm$^{-2}$, respectively). When fitting the spectra of these objects with MYTorus in decoupled mode, we find global column densities ($N_{\rm H,S}$) of $3.7^{+4.1}_{-2.6}\times10^{23}$ cm$^{-2}$ and $<5.5\times10^{23}$ cm$^{-2}$ for F2M 0830+3759 and F2M 1227+3214, respectively. Though this global gas obscuration is physically distinct from the gas which attenuates the emission along the line-of-sight to the central engine, it plays a role in reprocessing the observed X-ray spectrum. This obscuration geometry is consistent with the red quasar paradigm: while a direct view of the central engine is not completely blocked (i.e., broad emission lines are visible in the infrared spectrum), gas with large column densities is present near the black hole.

\item F2M 0830+3759, as originally pointed out by \citet{piconcelli}, has soft excess X-ray emission below 1 keV, which is well accommodated by a model where 7\% of the intrinsic AGN continuum leaks through holes in a patchy obscuring medium and is then scattered into or directly enters our line-of-sight. 

\item While F2M 1227+3214 has a measured $E(B-V)$/$N_{\rm H}$ value largely consistent with that of our Galaxy, the dust to gas ratio in F2M 0830+3759 is lower than the Galactic standard, though on the upper end of the distribution reported in the \citet{maiolino} sample (Figure \ref{nh_ebv}). Since F2M 0830+3759 lives in a host galaxy with a morphology indicative of a recent major merger \citep{urrutia2008}, the dust that reddens the optical quasar emission can be distributed on galaxy-wide scales, while the X-ray obscuring gas is likely circumnuclear, and perhaps within the dust sublimation zone of the broad-line region \citep[e.g.,][]{risaliti2009,risaliti2010,risaliti2011,maiolino2010}. Hence the disagreement between the observed and Galactic dust to gas ratio is perhaps to be expected. 

\item F2M 0830+3759 and F2M 1227+3214 have observed X-ray to 6$\mu$m luminosities consistent with the empirical relations derived for local Seyfert galaxies \citep{lutz} and unobscured quasars \citep{fiore2009,stern2015}, unlike the Type 2 SDSS quasars \citep{lansbury1,lansbury2} and the Compton-thick AGN discovered in the {\it NuSTAR} survey of COSMOS \citep{civano}, where the observed X-ray luminosities are heavily diminished.  Thus, X-ray observations of luminous obscured quasars, such as the two sources presented here, present a unique opportunity to test the X-ray to mid-infrared relationship in a new regime.

\end{itemize}

Red quasars similar to the sources discussed here may represent a short-lived, yet critical phase, in the growth of black holes and subsequent evolution of their host galaxies. {\it NuSTAR} data combined with recent advances in X-ray modeling, provide an unprecedented opportunity to peer through the obscuration and unravel the physical complexities of these systems. In particular, X-ray models that are capable of estimating a patchy distribution, where the line-of-sight column density is independently disentangled from the global column density, is of particular relevance for accurately understanding this population of AGN. Currently, two X-ray models have this capability: the clumpy torus model from \citet{liu2014} and the decoupled mode of the MYTorus model, and only the latter one is publicly available. Future observations of more red quasars will be essential for determining whether many have larger column densities than indicated purely by line-of-sight obscuration, and how this three-dimensional information may be related to larger-scale host galaxy obscuration.

\acknowledgments We thank the referee for a careful reading of this manuscript and helpful comments. This work was supported under NASA contract No. NNG08FD60C, and made use of data from the {\it NuSTAR} mission, a project led by the California Institute of Technology, managed by the Jet Propulsion Laboratory, and funded by the National Aeronautics and Space Administration. We thank the {\it NuSTAR} Operations, Software and Calibration teams for support with the execution and analysis of these observations. This research has made use of the {\it NuSTAR} Data Analysis Software (NuSTARDAS) jointly developed by the ASI Science Data Center (ASDC, Italy) and the California Institute of Technology (USA). AR is supported by the Gruber Science Fellowship. EG acknowledges the generous support of the Cottrell College Award through the Research Corporation for Science Advancement. WNB acknowledges support from Caltech NuSTAR subcontract 44A-1092750 and NASA ADP grant NNX10AC99G. RCH acknowledges support from NASA through ADAP award NNX12AE38G, the National Science Foundation through grant nos. 1211096 and 1515364, a Sloan Research Fellowship, and a Dartmouth Class of 1962 Faculty Fellowship. CR acknowledges financial support from the CONICYT-Chile grants ''EMBIGGEN" Anillo ACT1101, FONDECYT 1141218, Basal-CATA PFB--06/2007. 

Facilities: \facility{NuSTAR},\facility{CXC},\facility{XMM}

\end{document}